\documentclass[11pt,a4paper]{article}
\pdfoutput=1
\usepackage{jheppub}
\usepackage{rotating}
\usepackage{array}
\usepackage{amsmath}
\usepackage[normalem]{ulem}
\usepackage{slashed}
\usepackage{booktabs}
\usepackage[pdftex,table,dvipsnames]{xcolor}
\usepackage{units}
\usepackage{multirow}
\usepackage{url}

\usepackage{xspace}
\usepackage{subfig}
\usepackage{color}

\newcommand{\ga}{g_{a\gamma\gamma}}

\def \belletwo {Belle\,II\xspace}
\def \belle {Belle\xspace}
\def \babar {BaBar\xspace}

\def \superkekb {SuperKEKB\xspace}

\preprint{DESY-17-127}

\title{Revised constraints and \belletwo sensitivity for visible and invisible axion-like particles}

\author[1]{Matthew J. Dolan,}
\author[2]{Torben Ferber,}
\author[2,3]{Christopher Hearty,}
\author[4,5]{Felix Kahlhoefer}
\author[4]{and Kai Schmidt-Hoberg}

\affiliation[1]{ARC Centre of Excellence for Particle Physics at the Terascale, School of Physics, University of Melbourne, 3010, Australia}
\affiliation[2]{Department of Physics and Astronomy, University of British Columbia, Vancouver, British Columbia, V6T 1Z1 Canada}
\affiliation[3]{Institute of Particle Physics, Vancouver, British Columbia, V6T 1Z1 Canada}
\affiliation[4]{DESY, Notkestrasse 85, D-22607 Hamburg, Germany}
\affiliation[5]{Institute for Theoretical Particle Physics and Cosmology (TTK), RWTH Aachen University, D-52056 Aachen, Germany}

\emailAdd{dolan@unimelb.edu.au}
\emailAdd{ferber@physics.ubc.ca}
\emailAdd{hearty@physics.ubc.ca}
\emailAdd{felix.kahlhoefer@desy.de}
\emailAdd{kai.schmidt.hoberg@desy.de}

\abstract{Light pseudoscalars interacting pre-dominantly with Standard Model gauge bosons (so-called axion-like particles or ALPs) occur frequently in extensions of the Standard Model. In this work we review and update existing constraints on ALPs in the keV to GeV mass region from colliders, beam dump experiments and astrophysics. We furthermore provide a detailed calculation of the expected sensitivity of \belletwo, which can search for visibly and invisibly decaying ALPs, as well as long-lived ALPs. The \belletwo sensitivity is found to be substantially better than previously estimated, covering wide ranges of relevant parameter space. In particular, \belletwo can explore an interesting class of dark matter models, in which ALPs mediate the interactions between the Standard Model and dark matter. In these models, the relic abundance can be set via resonant freeze-out, leading to a highly predictive scenario consistent with all existing constraints but testable with single-photon searches at \belletwo in the near future.}

\keywords{Mostly Weak Interactions: Beyond Standard Model; Collider Physics: $e^+$-$e^-$ Experiments; Astroparticles: Cosmology of Theories beyond the SM}

\begin{document}

\maketitle

\flushbottom

\section{Introduction}

Axions and axion-like particles (ALPs) are a generic feature of many extensions of the Standard Model (SM), occurring for example in most solutions of the strong CP problem (see refs.~\cite{Hook:2014cda,Fukuda:2015ana} for recent work in this direction), in string compactifications~\cite{Arvanitaki:2009fg,Cicoli:2012sz} and in models with broken supersymmetry (the so-called R-axion~\cite{Bellazzini:2017neg}). Being Pseudo-Goldstone bosons, they can naturally be light and very weakly coupled, thus evading many of the strong constraints on new physics imposed for example by the LHC. In fact, the most promising way to search for ALPs may be the intensity frontier~\cite{Hewett:2012ns,Essig:2013lka}, i.e.\ searches with relatively low energy but very large integrated luminosity/intensity. The coming years promise significant progress in this area, driven first of all by the upcoming \belletwo experiment~\cite{Abe:2010gxa}, but also for example by NA62~\cite{NA62:2017rwk} and by longer-term projects such as the planned SHiP facility~\cite{Anelli:2015pba}. These searches may soon shed light on the existence and nature of ALPs.

ALPs with masses below the MeV scale can have a wide range of implications for cosmology and astrophysics~\cite{Cadamuro:2011fd}, affecting for example Big Bang Nucleosynthesis (BBN)~\cite{Millea:2015qra}, the Cosmic Microwave Background (CMB) and the evolution of stars. ALPs can constitute cold dark matter (DM)~\cite{Arias:2012az} and have also been considered as possible explanations for a range of astrophysical anomalies, such as the over-efficient cooling of certain classes of stars~\cite{Giannotti:2015kwo}, the surprising transparency of the Universe to very high-energy $\gamma$-rays~\cite{Meyer:2013pny} or the hints for a mono-energetic x-ray line around 3.5 keV~\cite{Cicoli:2014bfa,Conlon:2014xsa,Jaeckel:2014qea}.

Over the past few years an increasing amount of interest has been paid to ALPs in the MeV to GeV range~\cite{Mimasu:2014nea,Dolan:2014ska,Jaeckel:2015jla,Dobrich:2015jyk,Izaguirre:2016dfi,Knapen:2016moh,Brivio:2017ije,Bauer:2017nlg,Choi:2017gpf,Bauer:2017ris}. In this mass range, ALPs are largely irrelevant for astrophysics and cosmology, but they can have a number of interesting implications for particle physics. For example, ALPs have been considered as an explanation for the anomalous magnetic moment of the muon~\cite{Chang:2000ii,Marciano:2016yhf,Bauer:2017nlg} or for exotic resonances in nuclear transitions~\cite{Ellwanger:2016wfe}. Moreover, it has been pointed out that ALPs may play a crucial role in electroweak symmetry breaking and the solution of the hierarchy problem~\cite{Flacke:2016szy} via the so-called relaxion mechanism~\cite{Graham:2015cka}. Finally, ALPs offer an interesting possibility to connect the SM to a potential DM particle in such a way that thermal freeze-out can be reconciled with constraints from direct detection experiments~\cite{Nomura:2008ru,Boehm:2014hva,Dolan:2014ska}.

Given the level of detail and sophistication of many recent analyses, it is rather surprising that many constraints on the ALP parameter space have not been updated for a very long time. In fact, a number of constraints shown in recent studies are taken directly from the very early work by Mass\'{o} and Toldr\`{a}~\cite{Masso:1995tw,Masso:1997ru}, even though a wealth of newer experimental data has since become available. Moreover, the early calculations were never meant to be more than order-of-magnitude estimates and are therefore potentially misleading when compared to the projected sensitivities of future experiments. The aim of this paper is to revisit constraints on couplings of ALPs to photons and hypercharge gauge bosons for ALP masses in the MeV to GeV range and to compare existing constraints to the sensitivity of upcoming searches for ALPs. For this purpose, we will discuss bounds from electron-positron colliders, from electron beam dumps and from Supernova (SN) 1987A.\footnote{Constraints on ALPs from proton beam dump experiments have been studied in detail in ref.~\cite{Dobrich:2015jyk}.} While many of these constraints have been investigated in detail in the context of different models, such as hidden photons or light scalars (see e.g.~\cite{Batell:2009di,Andreas:2012mt,Essig:2013vha,Izaguirre:2013uxa,Batell:2014mga,Krnjaic:2015mbs}), we present the first comprehensive reinterpretation of these constraints in the context of ALPs.

A central part of our work is to calculate in detail the projected sensitivity to ALPs of \belletwo. While the primary purpose of \belletwo is to study the properties of B-mesons, the experiment is in fact ideally suited for a wide range of new-physics searches. Several previous studies have investigated the potential of \belletwo to search for invisible Dark Photon decays~\cite{Essig:2013vha,Izaguirre:2013uxa,lit:b2tip}, which will significantly extend the range of earlier searches at \babar~\cite{Aubert:2008as,delAmoSanchez:2010ac,Lees:2017lec}. Here we present the first realistic study of experimental backgrounds and detection efficiencies for ALP searches at \belletwo. We point out that previous studies have significantly underestimated the sensitivity of \belletwo for single-photon searches due to overly conservative background estimates based on an extrapolation of \babar results. Our study demonstrates that \belletwo can already explore new parameter space with early data and in the long run will be highly complementary to future searches for ALPs at the LHC~\cite{Bauer:2017ris} or at SHiP~\cite{Alekhin:2015byh,Dobrich:2015jyk}. We conclude that \belletwo therefore possesses a unique opportunity to discover both visibly and invisibly decaying ALPs.

The latter case is of particular interest in the context of models where the ALP is responsible for mediating the interactions between DM and SM particles. We focus on the case that DM couples pre-dominantly to photons. Strong constraints on this scenario from the CMB and indirect detection experiments can be avoided if DM annihilations during freeze-out are resonantly enhanced. We find that in this scenario the observed DM relic abundance can be reproduced in a well-defined region of parameter space, which is presently allowed by all experimental constraints but can be largely probed in the next few years with single-photon searches at \belletwo.

This paper is structured as follows. Section~\ref{sec:ALPs} provides a general review of the effective interactions of ALPs and establishes the notation used in the rest of this paper. The focus of section~\ref{sec:review} is on the discussion of existing constraints on the ALP parameter space. The special case of an ALP coupled to DM is discussed in section~\ref{sec:DM}. Finally, we present the projected sensitivity of \belletwo in section~\ref{sec:BelleII}, before concluding in section~\ref{sec:conclusions}.

\section{Effective interactions of ALPs}
\label{sec:ALPs}

This work focuses on the interactions of a pseudoscalar ALP $a$ with SM gauge bosons. Specifically, we consider the Lagrangian
\begin{equation}
\mathcal{L}= \frac{1}{2} \partial^\mu a \, \partial_\mu a - \frac{1}{2}m^{2}_{a} \, a^2 - \frac{c_B}{4 \, f_a} \, a \, B^{\mu\nu}\tilde{B}_{\mu\nu} - \frac{c_W}{4 \, f_a} \, a \, W^{i,\mu\nu}\tilde{W}^i_{\mu\nu} \; ,
\label{eq:L}
\end{equation}
where $B^{\mu\nu}$ and $W^{i,\mu\nu}$ denote the field strength of $U(1)_Y$ and $SU(2)_L$, respectively, and we have defined the dual field strength tensors via $\tilde{B}_{\mu\nu} = \frac{1}{2} \epsilon_{\mu\nu\rho\sigma} \, B^{\rho\sigma}$. The parameters $m_a$ and $f_a$ denote the ALP mass and decay constant, which we assume to be independent parameters.

We emphasize that the Lagrangian that we consider does not include all terms that would be expected to be present in a general effective field theory description of ALPs~\cite{Brivio:2017ije,Bauer:2017ris}. In particular, we do not consider interactions between ALPs and SM fermions or interactions between ALPs and gluons. This restriction is well-motivated in models where the interactions between ALPs and the SM arise from new heavy fermions that do not carry colour charge. The reason we are interested in ALPs that do not couple to gluons and fermions is that such interactions typically lead to flavour-changing processes (via penguin diagrams or via mixing with the $\pi^0$), which are tightly constrained by searches for rare decays~\cite{Dolan:2014ska,Choi:2017gpf}. ALPs coupling dominantly to the gauge bosons of $U(1)_Y$ and $SU(2)_L$, on the other hand, are much harder to probe experimentally and require dedicated experimental search strategies.\footnote{We note that from the interactions that we consider couplings to fermions will be introduced at the one-loop level and couplings to gluons will appear at the two-loop level. These couplings are however too small to have relevant experimental consequences~\cite{Bauer:2017ris}.}

Many of the constraints that we discuss below will remain valid even if the ALP has additional interactions. In fact, additional interactions are expected to increase the ALP production cross section and hence lead to even stronger bounds. Nevertheless, in some cases the presence of additional interactions may actually weaken the constraints. For example, additional decay modes will \emph{decrease} the ALP lifetime and therefore potentially suppress constraints from experiments searching for long-lived particles. Additional interactions may also lead to the trapping of ALPs in astrophysical objects, weakening the constraints obtained from such systems. The reader should therefore be careful when applying the bounds presented in this work to more complicated models.

After electroweak symmetry breaking, the two terms in eq.~(\ref{eq:L}) induce four different interactions between the ALP and SM gauge bosons:
\begin{equation}
 \mathcal{L} \supset - \frac{\ga}{4} a F_{\mu\nu} \tilde{F}^{\mu\nu} - \frac{g_{a\gamma Z}}{4} a F_{\mu\nu} \tilde{Z}^{\mu\nu} - \frac{g_{aZZ}}{4} a Z_{\mu\nu} \tilde{Z}^{\mu\nu} - \frac{g_{aWW}}{4} a W_{\mu\nu} \tilde{W}^{\mu\nu} \; ,
\end{equation}
where the field strengths and their duals are defined as above. The individual couplings can be calculated in terms of the parameters introduced above. The two couplings of greatest interest for the purpose of this work are
\begin{equation}
 \ga = \frac{c_B \, \cos^2 \theta_\mathrm{W} + c_W \, \sin^2 \theta_\mathrm{W}}{f_a} \, , \qquad  g_{a\gamma Z} = \frac{\sin 2 \theta_\mathrm{W} (c_W - c_B)}{f_a}  \; , 
\end{equation}
where $\theta_\mathrm{W}$ denotes the Weinberg angle.

If $c_B$ and $c_W$ are independent parameters, so are $\ga$ and $g_{a\gamma Z}$. In particular, for $c_B \approx c_W$ one finds $g_{a\gamma Z} \ll \ga$. Nevertheless, as pointed out in ref.~\cite{Izaguirre:2016dfi}, there are potentially strong constraints on $c_W$ from loop-induced flavour-changing processes like $B \to K a$. It is therefore particularly interesting to consider the case where $c_W \ll c_B$ and hence
\begin{equation}
 \ga \approx - \frac{1}{2} \cot \theta_\mathrm{W} g_{a\gamma Z} \approx - 0.94 g_{a\gamma Z} \; .
\end{equation}
We will refer to the case $c_B \sim c_W$ (and hence $g_{a\gamma Z} \ll \ga$) as \emph{photon coupling} and to the case $c_B \gg c_W$ (and hence $g_{a\gamma Z} \sim - \ga$) as \emph{hypercharge coupling}. We emphasize that in both cases ALPs will also couple to pairs of heavy gauge bosons, but we do not discuss the effect of these couplings further.

The $a\gamma\gamma$-interaction is of particular importance, as it determines the lifetime $\tau_a$ of the ALP. The decay width $\Gamma_a = \tau_a^{-1}$ is given by
\begin{equation}
\Gamma_a = \frac{\ga^2 \, m_{a}^3}{64 \pi} \; .
\end{equation}
It is worth emphasizing that for $\ga \ll 1\,\mathrm{TeV^{-1}}$ and for $m_a \ll 1\,\mathrm{GeV}$ this decay width is extremely small and hence the ALP decay length can be very large, in particular if the ALPs are produced with significant boost $\gamma_a = E_a / m_a$. For a detector of size $L_\mathrm{D}$ the fraction of ALPs that decay within the detector is given by\footnote{This expression assumes that either all ALPs are produced approximately in the same direction (as in beam dump experiments) or that the detector is approximately spherical. In all other cases both $L_\mathrm{D}$ and $p_a$ depend on the direction of the ALP momentum~\cite{Bauer:2017ris}.}
\begin{equation}
 p_a = 1 - \exp \left(- \frac{L_\mathrm{D}}{\gamma_a \, \tau_a} \right) \; .
\label{eq:decay}
\end{equation}
If the ALPs escape from the detector before decaying, this can prohibit searches for the decay $a \to \gamma\gamma$. On the other hand, such long decay lengths can facilitate a different kind of search, which focuses on the missing momentum carried away by the invisible ALP. We will discuss existing results and future prospects for both search strategies below.

\begin{figure}
\centering
\includegraphics[width=0.35\textwidth]{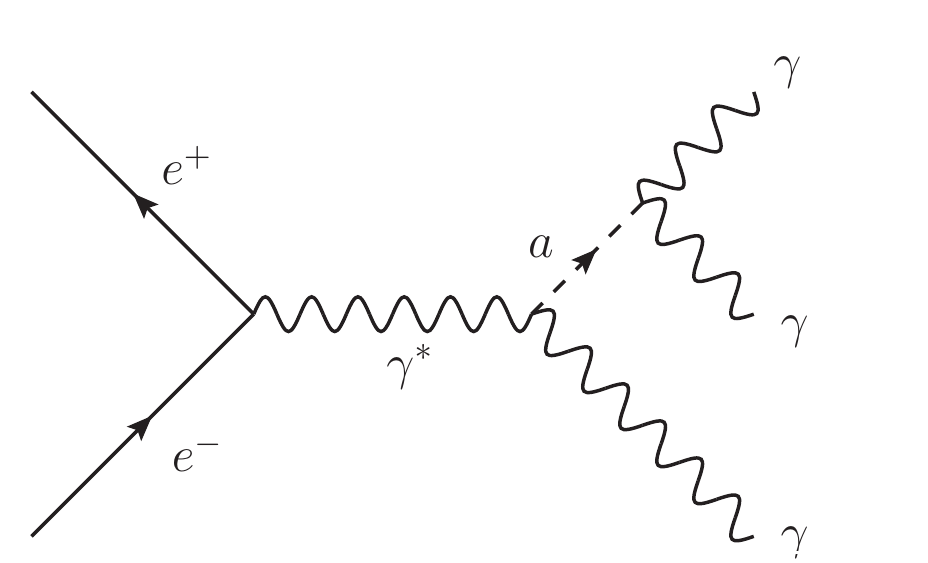}\qquad
\includegraphics[width=0.35\textwidth]{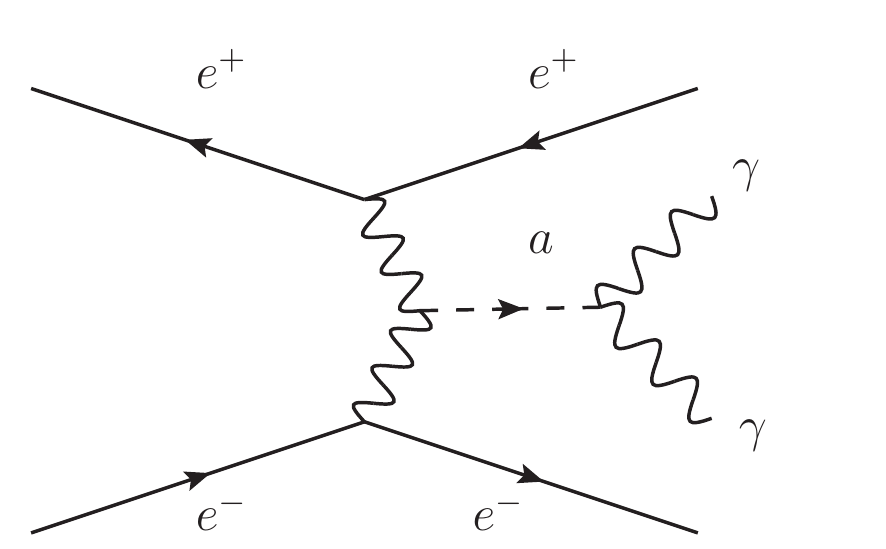}
\caption{\label{fig:production} Feynman diagrams for ALP production in $e^+ e^-$ collisions via ALP-strahlung (left) and photon fusion (right) and the subsequent decay of the ALP into two photons.}
\end{figure}

The same interaction can also be responsible for the production of ALPs, for example in $e^+ e^-$ collisions. There are two different production processes of interest: ALP-strahlung ($e^+ e^- \to \gamma^\ast \to \gamma + a$) and photon fusion ($e^+ e^- \to e^+ e^- + a$), see figure~\ref{fig:production}. For the former process (and in the limit $m_a \to 0$) the differential cross section with respect to the photon angle in the centre-of-mass (CM) frame is given by~\cite{Marciano:2016yhf}
\begin{equation}
 \frac{\mathrm{d}\sigma}{\mathrm{d}\cos\theta} = \frac{\ga^2 \, \alpha}{128} (3 + \cos 2 \theta) (1 - m_a^2/s)^3\; ,
\end{equation}
which has a mild angular dependence and is notably independent of the CM energy $\sqrt{s}$ for $m_a \ll \sqrt{s}$.\footnote{Even for very light ALPs there remains a slight dependence on $\sqrt{s}$ due to the running of both $\alpha$ and $\ga$, which can change by up to 10\% over the range of energies that we consider~\cite{Bauer:2017ris}.}
ALP-strahlung therefore typically leads to a photon with sizeable transverse momentum, which is a promising experimental signature. 

The cross section for ALP production via photon fusion can be calculated by replacing the colliding particles by their equivalent photon spectra $\gamma(x)$ and making use of the ALP production cross section from a pair of photons~\cite{Dobrich:2015jyk}:
\begin{equation}
 \sigma(\gamma \gamma \rightarrow a) = \frac{\pi\,\ga^2\,m_a}{16} \delta(m_{\gamma\gamma} - m_a) \; .
\end{equation}
Unless $m_a$ is close to $\sqrt{s}$, ALP production via photon fusion typically dominates over ALP-strahlung. However, the ALPs produced in this way are much harder to detect experimentally, as they carry only little energy and therefore decay into relatively soft photons in the laboratory frame. We will return to the experimental feasibility of searches for ALPs produced in photon fusion in section~\ref{sec:photonfusion}.

This work focuses on ALPs with mass below $10\,\mathrm{GeV}$, so that the decay $a \to \gamma Z$ is forbidden. The $a\gamma Z$ interaction nevertheless plays an important role, as it leads to the decay $Z \to \gamma + a$~\cite{Dror:2017nsg,Bauer:2017ris} with partial decay width given by
\begin{equation}
\Gamma(Z \to \gamma + a) = \frac{g_{a\gamma Z}^2}{384 \pi} \left(\frac{m_Z^2 - m_a^2}{m_Z}\right)^3\; .
\end{equation}
Depending on the ALP lifetime, this process can either lead to the signature $Z \to \gamma + \text{inv}$ or to $Z \to 3\gamma$, both of which can be tightly constrained by experiments.

To conclude this section we note that in principle ALPs may also be produced in Higgs decays, $h \to Z a$ or $h \to a a$, leading to strong constraints from the non-observation of these decay modes~\cite{Bauer:2017nlg,Bauer:2017ris}. These interactions however only appear when considering effective operators of dimension 6 or higher, and they are not directly linked to the interactions between ALPs and SM gauge bosons. While it is instructive to include these interactions in a general effective field theory approach, they are not generic and may be absent in specific UV completions. We will therefore not consider exotic Higgs decays in this work and instead focus on the phenomenology of the interactions between ALPs and gauge bosons.

\section{Review of bounds on the ALP parameter space}
\label{sec:review}

\begin{figure}
\centering
\includegraphics[width=0.45\textwidth]{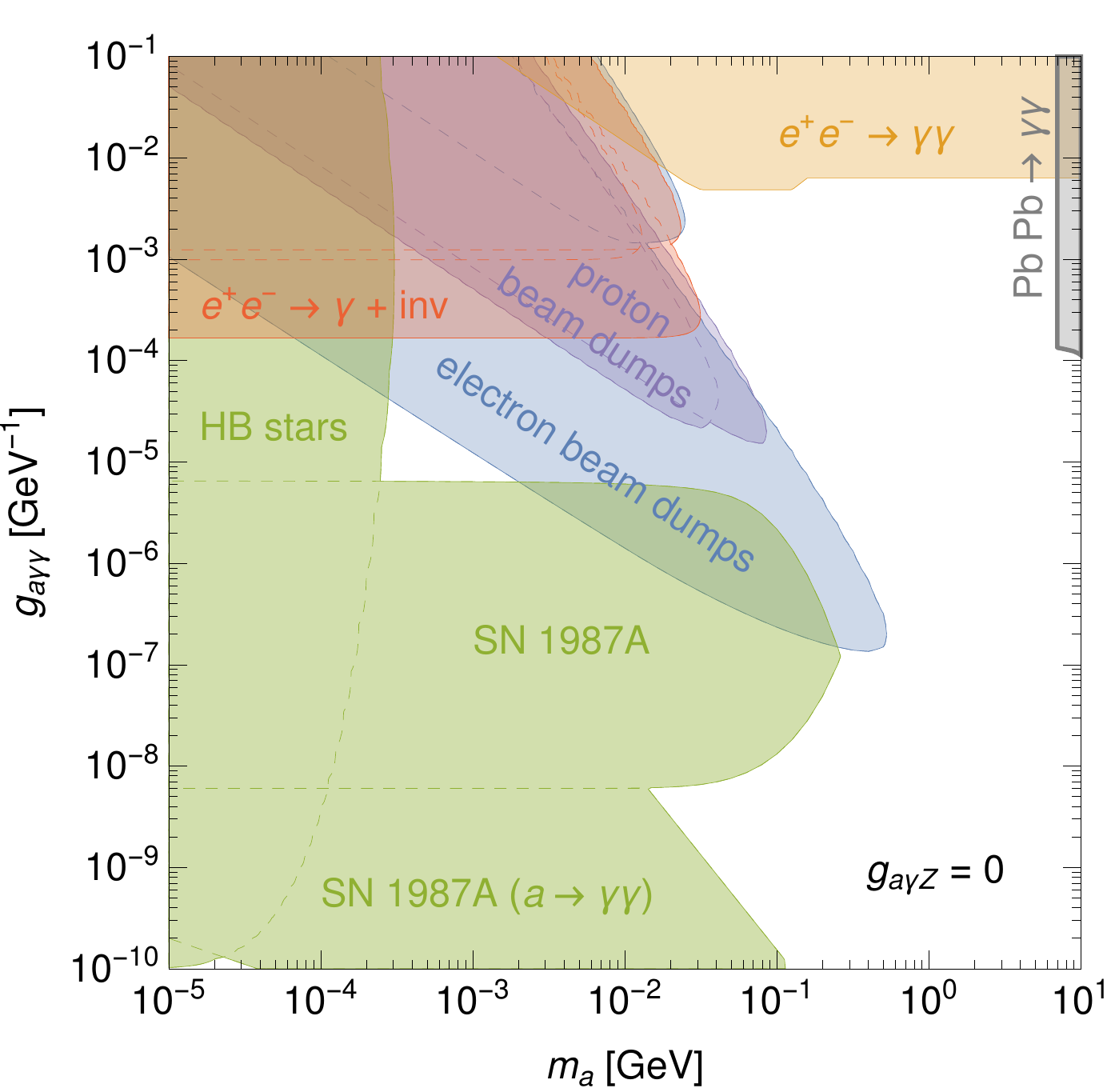}\qquad
\includegraphics[width=0.45\textwidth]{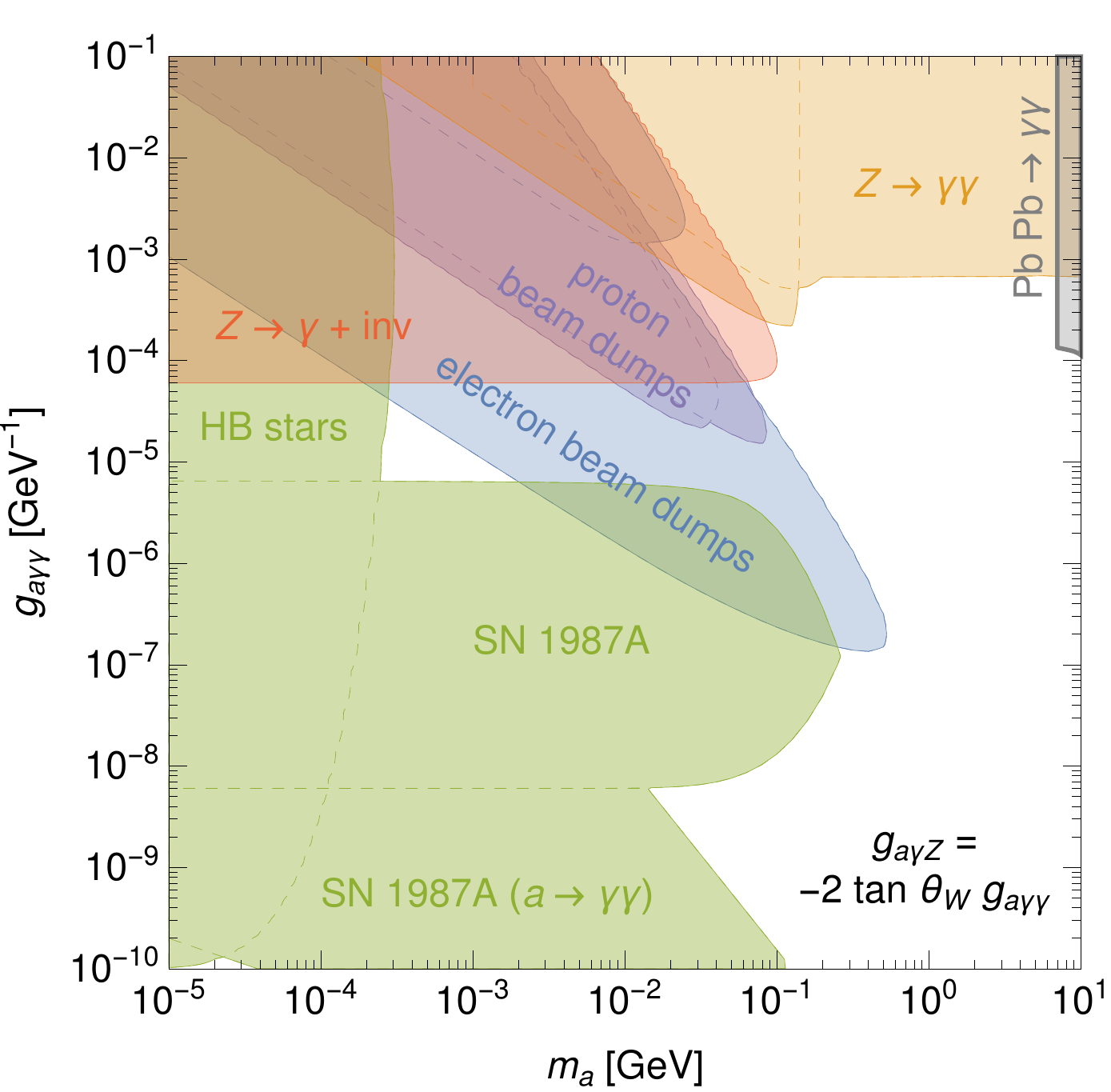}
\caption{\label{fig:constraints} Existing constraints on ALPs with photon coupling (left) and hypercharge coupling (right). Proton beam dump constraints are taken from ref.~\cite{Dobrich:2015jyk}, LEP constraints on $e^+ e^- \to \gamma \gamma$ from ref.~\cite{Jaeckel:2015jla}, CDF constraints on $Z \to \gamma\gamma$ from ref.~\cite{Bauer:2017ris}, bounds from horizontal branch stars from ref.~\cite{Cadamuro:2011fd}, bounds from visible decays of ALPs produced in SN 1987A from ref~\cite{Jaeckel:2017tud} and bounds from heavy-ion collisions from ref.~\cite{Knapen:2017ebd}. All other constraints have been revisited and updated in the present work.}
\end{figure}

In this section we review existing bounds on the ALP parameter space, updating constraints wherever new data or more precise calculations have become available. Most of the constraints that we will discuss only probe the effective ALP-photon coupling $\ga$. The only exception are constraints from high-energy colliders, which depend on whether the ALP couples to photons or hypercharge. We show a summary of all relevant constraints for both cases in figure~\ref{fig:constraints}. All collider and beam dump bounds are provided at $95\%$ confidence level (CL), with the exception of the bounds from \babar, which are provided at $90\%$ CL. Given that the parameter space under consideration covers many orders of magnitude, the difference between the two choices of CL is imperceptible.

\subsection{Bounds from electron-positron colliders}

\paragraph{Mono-photon searches at LEP.}

Relevant bounds on the ALP parameter space are obtained from so-called mono-photon searches, i.e.\ searches for highly-energetic photons in association with missing energy resulting from the process $e^+ e^- \to \gamma^\ast \to \gamma + a (\to \text{inv})$. The bound conventionally shown in the ALP parameter space is often attributed to LEP (see e.g.\ ref.~\cite{Mimasu:2014nea}), but actually goes back to the early analysis from ref.~\cite{Masso:1995tw} of a mono-photon search at the ASP experiment at SLAC~\cite{Hearty:1989pq}. Upon closer inspection, however, it turns out that this search is not sensitive to ALP masses in the sub-GeV region, because it requires $E_\gamma < 10\,\text{GeV}$ in the fiducial region, whereas for light ALPs the photon energy is given by $E_\gamma \approx \sqrt{s} / 2 = 14.5\,\mathrm{GeV}$.

To determine the sensitivity of LEP for ALPs, we follow the re-analysis from ref.~\cite{Fox:2011fx} of a mono-photon search at DELPHI~\cite{Abdallah:2008aa} based on $650\,\mathrm{pb^{-1}}$ at CM energies between 180 GeV and 209 GeV. The best sensitivity for an ALP signal stems from the High Density Projection Chamber, which covers $45^\circ < \theta < 135^\circ$, and from the three highest-energy bins included in the analysis, $0.9 \, E_\text{beam} < E_\gamma < 1.05 \, E_\text{beam}$. We implement detector efficiencies and resolution as detailed in ref.~\cite{Fox:2011fx} and assume that an ALP will escape unnoticed if it travels more than $L_\mathrm{D} = 260\,\mathrm{cm}$ in the radial direction without decaying~\cite{Aarnio:1990vx}.\footnote{The event selection includes a veto on energy depositions in any electromagnetic calorimeter. We hence approximate the detector length by the outer radius of the High Density Projection Chamber.}

We note that mono-photon searches have also been carried out at the LHC (for the most recent analyses see refs.~\cite{Aaboud:2017dor,CMS:2017ysu}), but their sensitivity does not significantly improve on the bound from LEP~\cite{Mimasu:2014nea}. Moreover, for these searches the validity of the ALP effective theory becomes a concern~\cite{Mimasu:2014nea,Brivio:2017ije}. We therefore do not show bounds from LHC mono-photon searches.

\paragraph{Radiative Upsilon decays.}
A related class of searches, which can be performed at $B$-factories, are searches for radiative decays of $\Upsilon(nS)$ with $n = 1,2,3$. While these searches are typically interpreted in terms of new invisible particles coupling to $b$-quarks, they also apply to the case where the new particle couples directly to the photon: $\Upsilon(nS) \to \gamma^\ast \to \gamma + a$. In fact, the corresponding branching ratio is easily calculated in terms of the branching ratio into electrons~\cite{Masso:1995tw}:
\begin{equation}
 \text{BR}(\Upsilon(nS) \to \gamma + a) = \frac{\alpha \, \ga^2 \, m_b^2}{8\pi} \cdot \text{BR}(\Upsilon(nS) \to e^+ e^-) \; ,
\end{equation}
where $m_b$ denotes the $b$-quark mass.

The bound conventionally shown comes from the Crystal Ball experiment, which gives $\text{BR}(\Upsilon(1S) \to \gamma + \text{inv}) < 4.0 \times 10^{-5}$~\cite{Antreasyan:1990cf}. However, much stronger bounds can be obtained from more recent measurements, such as the bound $\text{BR}(\Upsilon(3S) \to \gamma + \text{inv}) < 3 \times 10^{-6}$ from \babar~\cite{Aubert:2008as}. Here we reinterpret this latter constraint under the assumption that the ALP will escape from the detector, if it travels a distance of $L_\mathrm{D}=275$\,cm from the interaction point without decaying.\footnote{The event selection includes a veto of energy depositions in the instrumented flux return (IFR). We assume that a photon has sizeable detection efficiency only if it interacts before it reaches the two outermost IFR absorber layers.}

\paragraph{Dark Photon searches at \babar.}
\babar has published results from a search for invisibly decaying Dark Photons produced in association with an ordinary photon, $e^+ e^- \to \gamma A' (\to \text{inv})$~\cite{Lees:2017lec}. This search differs from an analogous ALP search in that the photon distribution peaks strongly at small polar angles $\theta$ in the CM frame. In the limit $m_{A'} \to 0$ one finds~\cite{Izaguirre:2013uxa}
\begin{equation}
\frac{\mathrm{d}\sigma}{\mathrm d \cos \theta} = \frac{(1 + \cos^2 \theta) \pi \, \alpha^2 \, \epsilon^2}{2 \, \sin^2 \theta \, E_\text{beam}^2} \; ,
\end{equation}
where $\epsilon$ denotes the kinetic mixing parameter.\footnote{Note that this expression at face value implies a divergent total cross section. In a more accurate treatment this is regulated by the finite electron mass, which we have neglected. Since realistic detectors will not be sensitive to photon angles close to $0$ or $\pi$, these details do not matter.} To convert a bound on $\epsilon$ for Dark Photons into a bound on $\ga$ for ALPs we therefore have to correct for the fact that the geometric acceptance will be very different in the two cases.

The \babar analysis considers $-0.6 < \cos \theta < 0.6$ for $m_{A'} > 5.5\,\text{GeV}$ and $-0.4 < \cos \theta < 0.6$ for $m_{A'} < 5.5\,\text{GeV}$. By integrating the respective differential cross sections for ALP production and Dark Photon production over these ranges we obtain the fiducial cross section including geometric acceptance. Using these numbers, we can translate bounds on Dark Photons into the ALP parameter space under the assumption that all other selection cuts have the same efficiency for the two models. For very small masses of the invisibly decaying particle, we find that the translation is given by
\begin{equation}
 \ga = 1.8 \times 10^{-4} \, \text{GeV}^{-1} \left(\frac{\epsilon}{10^{-3}}\right) \; . 
\end{equation}
Repeating this calculation for finite ALP masses and taking into account the probability that the ALP decays before leaving the detector (see above) using a detector length of $L_\mathrm{D}=275$\,cm \cite{Aubert:2001tu}, we can then reinterpret the full \babar bound in the context of ALPs.

\paragraph{Radiative Z-boson decays.}
If $g_{a\gamma Z}$ is non-zero, ALPs can also be produced in the decay $Z \to \gamma + a$. The resulting experimental signature depends on the ALP decay length. If the decay length is large compared to the size of the detector, the most promising search channel is $Z \to \gamma + \text{inv}$. This process has for example been studied by the L3 collaboration at LEP~\cite{Acciarri:1997im}, which quotes an upper limit on the corresponding branching ratio of $\text{BR}(Z \to \gamma + a) < 1.1 \times 10^{-6}$. We reinterpret this bound under the assumption that the ALP escapes if it does not decay within $L_\mathrm{D} = 180\,\mathrm{cm}$ from the interaction point~\cite{L3:1989aa}.\footnote{The event selection includes a veto of energy depositions in any calorimeter. We hence approximate the detector length by the outer radius of the barrel hadronic calorimeter.}

Constraints of similar strength are obtained for ALPs decaying close to the interaction point. If the ALP has a high boost but a short lifetime, the two photons produced in its decay will be approximately collinear and may thus appear in the detector as a single photon. The resulting signature can then mimic the forbidden decay $Z \to 2\gamma$~\cite{Jaeckel:2015jla,Chala:2015cev}. Constraints on this decay mode from LEP have been discussed in ref.~\cite{Jaeckel:2015jla}, but ref.~\cite{Bauer:2017ris} points out that for $m_a \lesssim m_{\pi^0}$ even stronger constraints can be derived from the CDF result $\text{BR}(Z \to \gamma \gamma) < 1.45 \times 10^{-5}$~\cite{Aaltonen:2013mfa}.

For $m_a \gtrsim 10\,\mathrm{GeV}$, on the other hand, the photons from the ALP decay can easily be distinguished and one obtains the signature $Z \to 3\gamma$. A recent ATLAS search constrains this decay mode to $\text{BR}(Z \to 3\gamma) < 2.2 \times 10^{-6}$~\cite{Aad:2015bua}. Future LHC searches are expected to significantly improve this bound and to extend the sensitivity to lower ALP masses~\cite{Bauer:2017ris}.

\subsection{Bounds from beam dump experiments}

ALPs can be produced in electron and proton beam dump experiments via Primakoff production, i.e.\ the conversion of a photon into an ALP in the vicinity of a nucleus~\cite{Tsai:1986tx}. A number of relevant electron beam dump experiments have been studied in ref.~\cite{Bjorken:2009mm}, while constraints from proton beam dump experiments have been reviewed in ref.~\cite{Dobrich:2015jyk}.

\paragraph{SLAC E141.}
Although E141 primarily searched for long-lived particles decaying to $e^+ e^-$~\cite{Riordan:1987aw}, during some of the data taking a photon converter was inserted in front of the detector, giving the experiment sensitivity also for ALPs decaying to photons. The results from this search have been presented in ref.~\cite{Krasny,Dobrich:2017gcm}. Although results are only provided for a limited range in ALP masses, the given details on the detector geometry are sufficient to extend the search range by rescaling with the appropriate decay probability and assuming that the positron energy spectrum is independent of the ALP mass as long as $m_a \ll E_\text{beam}$.

To perform the rescaling of the decay probability, we need to make an assumption on the typical energy of ALPs that leads to an observable signal, i.e.\ a positron with $E > E_\text{beam} / 2$. We find that good agreement with ref.~\cite{Krasny} is obtained for $E_a \approx 6.5 \, \text{GeV}$. Previous attempts to reinterpret data from E141 have assumed that smaller ALP energies are sufficient to produce sufficiently highly-energetic positrons, leading to somewhat more aggressive bounds~\cite{Hewett:2012ns}.

\paragraph{SLAC E137.}
The E137 experiment at SLAC has published results from a dedicated search for ALPs coupling only to photons~\cite{Bjorken:1988as}. The paper does not consider the turnover of the exclusion limit towards large couplings due to the exponential suppression of the number of ALPs that reach the detector, but it is possible to include this additional effect with the information provided in the paper. Specifically, we start from the photon track-length distribution as a function of energy provided in figure~14 of ref.~\cite{Bjorken:1988as} and assume that the cooling water in the beam dump yields the dominant contribution to the ALP production.\footnote{Assuming the dominant production to come from the aluminium plates does not significantly change our results. While the production rate for aluminium is larger than that for oxygen by around 20\%, this is compensated by aluminium having a slightly larger form factor suppression.}

Although the total decay length between absorber and detector is $204\,\mathrm{m}$, ALPs decaying at the beginning of the decay volume will often not lead to an observable signal, since the photons produced in the decay may miss the detector, which only has a radius $r_\mathrm{D} \approx 1.5\,\mathrm{m}$. The typical opening angle between the two photons produced in the ALP decay is $\theta_{\gamma\gamma} \sim 2 / \gamma_a = 2 \, m_a / E_a$. If the ALP decay happens at a distance $\Delta z$ from the detector, the two photons will thus hit the detector with a typical separation of $\theta_{\gamma\gamma} \Delta z$. To be conservative, we therefore assume that an ALP decay will only lead to an observable signal in the detector if $\Delta z < E_a \, r_\mathrm{D} / m_a$, so that the effective length of the decay volume is reduced for ALPs with small boost factor. This procedure allows us to extend the bound from ref~\cite{Bjorken:1988as} to the full parameter space relevant for ALP searches. Our final result roughly agrees with the one shown in ref.~\cite{Hewett:2012ns}, although we find slightly stronger constraints towards larger couplings by using a more realistic photon distribution in our calculation.

\paragraph{Proton beam dumps.}
Constraints on the ALP parameter space from proton beam dumps have recently been studied in ref.~\cite{Dobrich:2015jyk}. Here we show the bounds obtained in that work from CHARM~\cite{Bergsma:1985qz} and NuCal~\cite{Blumlein:1990ay,Blumlein:1991xh}.

\subsection{Bounds from heavy-ion collisions}

Ref.~\cite{Knapen:2016moh} proposed a novel way to search for ALPs in ultraperipheral heavy-ion collisions. In such collisions ALPs can be produced via the fusion of two coherently emitted photons, i.e.\ via diagrams analogous to the one shown in the right panel of figure~\ref{fig:production}. The resulting ALPs have very low boost factors and their decays consequently lead to two very soft photons that are approximately back-to-back. The minimum photon energy required by the trigger places a lower bound on the ALP masses that can be probed with this search strategy. A reinterpretation of the recent ATLAS search for light-by-light scattering~\cite{Aaboud:2017bwk} leads to relevant constraints for ALPs with $m_a > 7\,\mathrm{GeV}$~\cite{Knapen:2017ebd}.

\subsection{Bounds from astrophysics}

\paragraph{Supernova 1987A.}
Weakly coupled particles such as axions or ALPs with masses up to about $100\,\text{MeV}$ can be copiously produced in the hot core of a supernova. Because of their weak couplings these particles stream out of the core and thereby constitute a new energy loss mechanism. In the absence of such new particles the main cooling mechanism is due to neutrino emission. The corresponding neutrino signal has been observed in the case of SN 1987A, placing a bound on possible exotic energy loss mechanisms, which should not exceed the energy loss via neutrino emission.

ALPs that couple exclusively to photons are produced via the Primakoff process. As electrons are highly degenerate in the supernova core, their phase space is Pauli-blocked and their contribution to ALP production is negligible. Protons are only partially degenerate and correspondingly the process $\gamma + p \rightarrow p + a$ is the main production mode. The resulting ALP energy spectrum has been calculated with detailed account of the production process in a core-collapse supernova~\cite{Payez:2014xsa}. An analytical fit of these results has been performed in ref.~\cite{Jaeckel:2017tud}, which yields an approximate expression for the ALP production rate:
\begin{equation}
\frac{\mathrm{d}N_{a}}{\mathrm{d}E_a} \sim \frac{C \, E_a^2 \, \sigma_\text{pr}(E_a)}{\exp(E_a/T)-1} \; ,
\end{equation}
where $T$ is the effective temperature and $\sigma_\text{pr}(E_a)$ is the cross section for Primakoff production~\cite{Cadamuro:2011fd}. For SN 1987A a good fit is obtained for $C = 2.54 \times 10^{77} \, \mathrm{MeV^{-1}}$ and $T = 30.6 \,\mathrm{MeV}$. The total energy outflow is then given by 
\begin{equation}
E = \int_{m_a}^{\infty} \frac{\mathrm{d}N_{a}}{\mathrm{d}E_a} E_a \,\mathrm{d}E_a \; ,
\end{equation}
which should be smaller than $ \sim 3 \times 10^{53}\,\text{erg}$. Using this constraint leads to an upper bound on the ALP coupling of around $\ga < 6 \times 10^{-9} \, \text{GeV}^{-1}$ for small ALP masses.

This bound however does not apply for arbitrarily large couplings, because at some point the axions will interact so strongly that they are trapped in the supernova core. The ALP mean free path is given by
\begin{equation}
\lambda_a=\frac{1}{n_p \, \sigma_\text{bc}} \; ,
\end{equation}
where $n_p$ denotes the proton density and $\sigma_\text{bc}$ is the cross section for back-conversion, which is directly related to the production cross section~\cite{Brockway:1996yr}:
\begin{equation}
 \sigma_\text{bc} \simeq \frac{2}{\beta_a^2} \sigma_\text{pr}(E_a)
\end{equation}
with $\beta_a = \sqrt{1 - m_a^2/E_a^2}$. But even if the ALP mean free path is smaller than the size of the supernova core ($\sim 10\,\text{km}$), energy transport via ALPs can still be large. The new particle is harmless only if it interacts more strongly than the particles which provide the standard mode of energy transfer, i.e.\ neutrinos.

To estimate the rate of energy transport, we can calculate the ALP Rosseland mean opacity $\kappa_a$. For the inverse Primakoff process this can be written as~\cite{Raffelt:1996wa}
\begin{equation}
\kappa_{a}^\text{P} = \frac{\int_{m_a}^{\infty}  \mathrm{d}E_a \, E_a^3 \, \beta_a \, \frac{\partial}{\partial T}\frac{1}{\exp(E_a/T)-1}}{ \rho \int_{m_a}^{\infty}  \mathrm{d}E_a \, \lambda_a \, E_a^3 \, \beta_a  \frac{\partial}{\partial T}\frac{1}{\exp(E_a/T)-1} } \; ,
\end{equation}
while the contribution from decay is given by~\cite{Raffelt:1988rx}
\begin{equation}
\kappa_a^\text{D} = \frac{(2\pi)^{7/2}}{45 \, \rho}\left(\frac{T}{m_a}\right)^{5/2} e^{m_a/T} \, \Gamma_{a\gamma\gamma} \; .
\end{equation}
The constraint from SN 1987A only applies if $\kappa_a=\kappa_a^\text{P}+\kappa_a^\text{D} < \kappa_\nu$, where $\kappa_\nu \approx 8 \times 10^{-17} \mathrm{cm^2/g}$ is the neutrino opacity~\cite{Masso:1995tw}. For small ALP masses this corresponds to $\ga \lesssim 7 \times 10^{-6}\,\mathrm{GeV^{-1}}$.

An additional constraint from SN 1987A can be obtained by considering not only the energy loss due to ALP emission, but also the visible signal resulting from the ``ALP burst'' if the ALPs subsequently decay into photons~\cite{Jaeckel:2017tud}. In figure~\ref{fig:constraints} we show the constraint obtained in ref.~\cite{Jaeckel:2017tud}, which extends the bound from SN 1987A to even smaller couplings.

We emphasize that our treatment of the limit from SN1987A remains somewhat simplistic. Our parametrisation of the ALP energy spectrum and outflow are based on a fit to simulations performed with $\ga=10^{-10}\mathrm{GeV^{-1}}$ for a light ALP, and it is unclear how accurate this is over for large values $\ga$ and for $m_a$ near the supernova core temperature. Modern models of supernovae (see ref.~\cite{Janka:2012wk}) are also considerably more sophisticated than ours, which assumes a fixed nuclear density and temperature (following the treatment in refs.~\cite{Raffelt:1987yb,Masso:1995tw}). The mean free path of neutrinos depends sensitively on radius, time and energy, making it difficult to encode the effect in terms of a single opacity~\cite{Buras:2005rp}. We estimate this introduces a theoretical uncertainty of at least a factor of two into our limits. To reduce this uncertainty will likely require a dedicated simulation of heat transfer in the core of a supernova. While some work in this direction has been undertaken for axions~\cite{Payez:2014xsa,Fischer:2016cyd}, it would be interesting to perform dedicated simulations for the ALP scenario.

\paragraph{Horizontal branch stars.}
Constraints on the ALP parameter space can be obtained by considering the cooling of horizontal branch (HB) stars~\cite{Raffelt:1987yb,Raffelt:1996wa}. These constraints have recently been investigated in ref.~\cite{Cadamuro:2011fd}, and we show the bounds derived there. As with supernovae, it would be interesting to undertake more detailed simulations for ALPs, as has been done for axions in ref.~\cite{Friedland:2012hj}.

\paragraph{Big Bang Nucleosynthesis.}
It was pointed out in ref.~\cite{Cadamuro:2011fd} that the ALP parameter space is potentially strongly constrained by BBN, if the ALP decay rate is comparable to the Hubble rate during BBN. In this case, ALP decays into photons may alter the baryon-to-photon ratio and the number of relativistic degrees of freedom, leading in particular to conflict with the well-measured ratio of the abundances of deuterium and helium. These constraints were updated in ref.~\cite{Millea:2015qra} and shown to strongly disfavour the triangular region between the bounds from beam dump experiments and the constraints from SN 1987A and HB stars. Nevertheless, these constraints rely on the assumptions of a standard cosmological history and can be significantly weakened for example if additional relativistic degrees of freedom are present during BBN, making them more model-dependent than the other constraints considered here. We will therefore not show the BBN constraints in the following.

\section{ALPs coupled to dark matter}
\label{sec:DM}

In this section we will extend the model discussed above and consider the case of ALPs coupled to DM particles. The immediate consequence of such a coupling is that, provided the DM particle is sufficiently light, the ALP obtains an invisible decay mode. This will enhance the sensitivity of searches based on missing energy and suppress the sensitivity of searches that rely on the reconstruction of a visible final state.

For concreteness let us assume that the DM particle is a Majorana fermion $\chi$. The generic interaction with ALPs is then of the form
\begin{equation}
 \mathcal{L}_\text{DM} = g_{a\chi\chi} \, \bar{\chi} \gamma^\mu \gamma^5 \chi \, \partial_\mu a\; ,
\end{equation}
where $g_{a\chi\chi}$ has mass dimension $-1$, just like $\ga$. The invisible decay width is then given by
\begin{equation}
\Gamma_\text{inv} = \frac{g_{a\chi\chi}^2 \, m_a \, m_\chi^2}{\pi} \sqrt{1 - \frac{4 \, m_\chi^2}{m_a^2}} \; , \label{eq:Gammainv}
\end{equation}
where $m_\chi$ denotes the DM mass and $\Gamma_\text{inv} = 0$ for $m_\chi > m_a / 2$. If the coupling $g_{a\chi\chi}$ is large compared to $\ga$ and $m_\chi$ is not much lighter than $m_a$, the invisible decay width can easily become large relative to the visible one.\footnote{This is for example the case if we assume that $\ga$ is generated at loop-level, whereas $g_{a\chi\chi}$ is generated at tree-level. It is then very plausible that $g_{a\chi\chi}$ is several orders of magnitude larger than $\ga$.}

Let us assume that this is indeed the case and that $\text{BR}(a \to \chi\chi) \approx 100\%$. Relevant experimental constraints then come from mono-photon searches at LEP and \babar (as well as from LEP bounds on $Z \to \gamma + \text{inv}$ if the ALP couples to hypercharge). The constraints from SN 1987A will also become much stronger, as the DM particles produced in ALP decays can contribute to supernova cooling even in parameter regions where the ALP decay length and mean free path are small.\footnote{
In principle even DM particles could be trapped inside the supernova, if their interaction rate with real photons ($\chi \gamma \to \chi \gamma$) and with virtual photons ($\chi p \to \chi p + \gamma$) is large enough. The rate of these processes is however suppressed relative to the rate of ALP back-conversion by a factor of $g_{a\chi\chi}^2 \, m_\chi^2 \ll 1$. We find that the DM mean free path is large compared to the size of the supernova as long as $m_\chi \, \ga \, g_{a\chi\chi} \lesssim 10^{-7} \, \mathrm{GeV^{-1}}$.}

Before we compare the sensitivity of the different searches, let us briefly discuss a related question: Is it possible within the simple model introduced above to reproduce the observed DM relic abundance while at the same time evading constraints from various DM searches? If the process $\chi \chi \to \gamma \gamma$ has a cross section close to the thermal one, $(\sigma v)_\text{th} \approx 3 \times 10^{-26} \mathrm{cm^3/s}$, the DM particle can in principle obtain its relic abundance from thermal freeze-out. For $m_\chi \ll m_a$ the annihilation cross section is given by
\begin{equation}
 \sigma(\chi \chi \to \gamma \gamma) v \simeq \frac{4 \, \ga^2 \, g_{a\chi\chi}^2 \, m_\chi^6}{\pi \, m_a^4} \; ,
\end{equation}
where $v$ denotes the relative velocity of the two DM particles in the CM frame. This cross section is however tiny even for optimistic parameter choices. For example, choosing $\ga = 10^{-3}\:\text{GeV}^{-1}$, $g_{a\chi\chi} = 10^{-2}\:\text{GeV}^{-1}$, $m_\chi = 1\:\text{GeV}$ and $m_a = 5\:\text{GeV}$, one obtains $\sigma v \approx 2 \times 10^{-30} \mathrm{cm^3 / s}$.

A second difficulty arises because the annihilation cross section is $s$-wave (velocity unsuppressed). Consequently, if $\sigma v = (\sigma v)_\text{th}$, the annihilation of DM into mono-energetic photons should still be observable in the present Universe. However, searches for $\gamma$-ray lines from Fermi-LAT exclude the thermal cross section for $s$-wave annihilation into photons across the entire energy range~\cite{Ackermann:2015lka}.\footnote{While the Fermi-LAT collaboration only considers DM masses down to $200\,\text{MeV}$, independent studies have extended these bounds down to $100\:\text{MeV}$~\cite{Albert:2014hwa}. We note in passing that constraints from direct detection experiments are strongly suppressed for this scenario~\cite{Frandsen:2012db} and can be neglected.} Thus, thermal freeze-out into photons is only viable if the annihilation cross section has a strong velocity dependence.

A simple way to both enhance the annihilation cross section and to introduce a strong velocity dependence is to consider the case that the DM mass is close to resonance \hbox{$m_\chi \approx m_a / 2$}. In this case, one needs to consider the full expression for the annihilation cross section:
\begin{equation}
\sigma(\chi \chi \to \gamma \gamma) v = \frac{\ga^2 \, g_{a\chi\chi}^2}{4 \pi} \frac{m_\chi^2 \, s^2}{(m_a^2 - s)^2 + 
  m_a^2 \, \Gamma_a^2} \; ,
  \label{eq:sigv}
\end{equation}
where $\sqrt{s}$ denotes the CM energy. The quantity relevant for the calculation of the relic abundance is then the thermally averaged annihilation cross section~\cite{Gondolo:1990dk}
\begin{equation}
\langle \sigma v \rangle = \int_{4 \, m_\chi^2}^\infty \frac{s \sqrt{s - 4 \, m_\chi^2} K_1(\sqrt{s} / T) \sigma v}{16 \, T \, m_\chi^4 \, K_2(m_\chi/T)^2} \mathrm{d}s \; , \label{eq:sigthermal}
\end{equation}
where $T$ denotes the temperature of the thermal bath and $K_i$ denote the spherical Bessel functions of the second kind.

\begin{figure}
\centering
\includegraphics[width=0.7\textwidth]{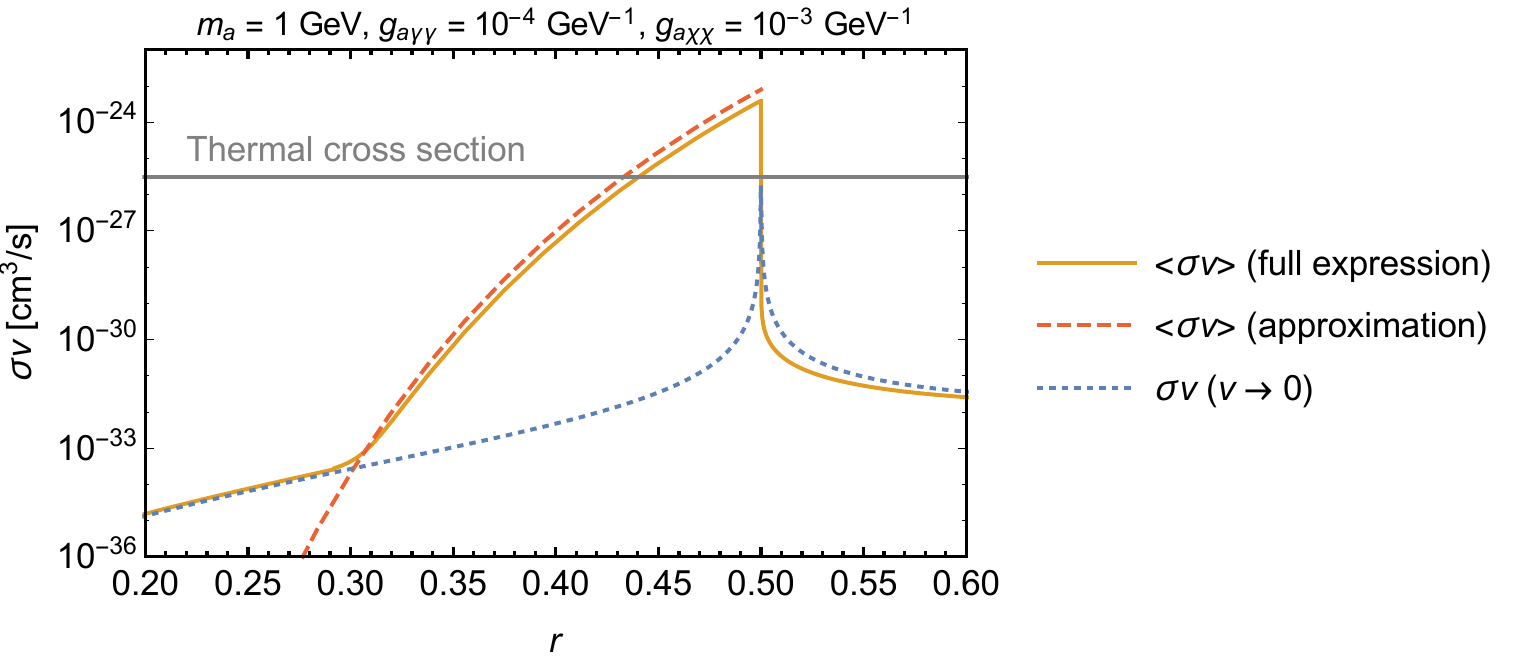}
\caption{\label{fig:sigmav} Comparison of the thermally averaged annihilation cross section $\langle \sigma v \rangle$ (orange, solid) and the annihilation cross section in the limit $v \to 0$ (blue, dotted) as a function of the mass ratio $r = m_\chi / m_a$ for $m_a = 1 \: \mathrm{GeV}$ and $\ga = 10^{-4}\:\mathrm{GeV}^{-1}$. The red dashed line indicates the simplified expression from eq.~(\ref{eq:sigsimple}), which is valid close to the resonance. Away from the resonance the cross section depends also on the ALP-DM coupling, which has been set to $g_{a\chi\chi} = 10^{-3}\:\mathrm{GeV}^{-1}$.}
\end{figure}

If $m_\chi$ is close to (but slightly below) $m_a / 2$, the integrand in eq.~(\ref{eq:sigthermal}) is strongly peaked around $s \approx m_a^2$, such that the intermediate ALP is on-shell and the annihilation receives a resonant enhancement. We can then make the replacement $s \to m_a^2$ everywhere except in the denominator of eq.~(\ref{eq:sigv}), in which case the integration can be performed analytically. If we furthermore substitute $\Gamma_\text{inv}$ from eq.~(\ref{eq:Gammainv}) for $\Gamma_a$, we obtain the very simple expression
\begin{equation}
\langle \sigma v \rangle \simeq \frac{\pi \, \ga^2 \, x \, K_1(x/r)}{64 \, r^5 \, K_2(x)^2} \; , \label{eq:sigsimple}
\end{equation}
where we have introduced the dimensionless temperature ratio $x = m_\chi / T$ and the dimensionless mass ratio $r = m_\chi / m_a$. We make two important observations
\begin{enumerate}
 \item Contrary to naive expectation, $\langle \sigma v \rangle$ does not depend on $g_{a\chi\chi}$. The reason is that increasing $g_{a\chi\chi}$ enhances the production cross section for the ALP in the intermediate state, but at the same time broadens the resonance. These two effects cancel exactly, so that only the coupling to photons enters in the final expression.
 \item Even though $\langle \sigma v \rangle$ is exponentially sensitive to the mass ratio $r$ (since $K_1(x/r) \propto \exp(-x/r) \sqrt{r}$), it becomes immediately clear that it is not necessary to tune $m_\chi$ exactly to $m_a / 2$. For example, the difference between $r = 0.49$ and $r = 0.499$ is only about a factor of 2 (assuming $x = 20$, which is a typical value of the temperature ratio during freeze-out).
\end{enumerate}
This latter point is illustrated in figure~\ref{fig:sigmav}, which shows in blue (dotted) the annihilation cross section without thermal averaging, in orange (solid) the annihilation cross section with thermal averaging and in red (dashed) the approximate expression given in eq.~(\ref{eq:sigsimple}).

As we have seen above, the thermally averaged annihilation cross section (and hence the DM relic abundance) depends only on $\ga$ and the mass ratio $r$. If we fix the mass ratio $r$ we can therefore determine the value of $\ga$ that yields the observed relic abundance.\footnote{When calculating the relic abundance by solving the Boltzmann equation, a weak additional dependence on the DM mass arises from the fact that the number of relativistic degrees of freedom and hence the expansion rate of the Universe depends slightly on the temperature during freeze-out. The value of $\ga$ implied by the observed relic abundance therefore depends slightly on $m_a$ even for fixed mass ratio $r$. We take this dependence into account by solving the Boltzmann equation numerically using \texttt{micrOmegas v4.2.5}~\cite{Belanger:2014vza}. We note that this calculation assumes that DM couples sufficiently strongly to photons that it remains in kinetic equilibrium during freeze-out~\cite{Binder:2017rgn}.} It turns out that the resulting values of $\ga$ are in an experimentally interesting region: For $r = 0.45$ ($r = 0.49$) we find roughly $\ga \sim 10^{-4}\:\mathrm{GeV^{-1}}$ ($\ga \sim 10^{-5}\:\mathrm{GeV^{-1}}$). These estimates indicate the regions of parameter space that are interesting for resonant thermal freeze-out.\footnote{If $r$ is extremely close to 0.5, the ALP-photon coupling can even be somewhat smaller than $\ga = 10^{-5}\:\mathrm{GeV^{-1}}$, but in this case indirect detection constraints again become relevant, as resonant enhancement of annihilation processes can also occur in the present Universe (see figure~\ref{fig:sigmav}).} In the following section we will compare these values to the sensitivity that can be achieved by \belletwo (see figure~\ref{fig:BelleII_inv}).

\section{Sensitivity to ALPs of \belletwo}
\label{sec:BelleII}

The \belletwo experiment at the SuperKEKB accelerator is a second generation \hbox{B-factory} and successor of the \belle and \babar experiments \cite{Abe:2010gxa}. It is currently under construction and will start data taking in 2018. \superkekb is a circular asymmetric $e^+e^-$ collider with a nominal collision energy of $\sqrt{s}$\,=\,10.58\,GeV. The design instantaneous luminosity is \hbox{$8\times10^{35}$\,cm$^{-2}$ s$^{-1}$}, which is about 40 times higher than at the predecessor collider KEKB. The \belletwo detector is a large-solid-angle magnetic spectrometer. Three  sub-detectors are particularly relevant for the ALP searches described in this paper: A 56-layer central drift chamber (CDC) is used for tracking of charged particles and covers a polar angle region of $(17-150)^{\circ}$. The electromagnetic calorimeter (ECL) comprising CsI(Tl) crystals with an upgraded waveform sampling readout for beam background suppression covers a polar angle region of $(12-155)^{\circ}$ and is located inside a superconducting solenoid coil that provides a 1.5\,T magnetic field. The ECL has inefficient gaps between the endcaps and the barrel for polar angles between $(31.3-32.2)^{\circ}$ and $(128.7-130.7)^{\circ}$. An iron flux-return is located outside of the magnet coil and is instrumented with resistive plate chambers and plastic scintillators to mainly detect $K^0_L$ mesons, neutrons, and muons (KLM) that covers a polar angle region of $(25-155)^{\circ}$.

We study the sensitivity for a dataset corresponding to an integrated luminosity of 20\,fb$^{-1}$, which is expected to be collected in 2018 without vertex detectors installed. We then scale the expected sensitivity $S(\ga)$  to the planned full integrated luminosity of 50\,ab$^{-1}$ after about 7\,years of running, using $S(\ga) \propto \sqrt[4]{\mathcal{L}}$. We argue that the expected increase of beam induced background rates at highest luminosity are not relevant for these searches.

In the following we consider ALP decays into DM and into two photons from ALPs produced in ALP-strahlung ($a\to\chi \chi$ and $a\to\gamma\gamma$) and photon-fusion production ($a\to\gamma\gamma$ only). 

\subsection{ALP decays into dark matter}

We study decays of ALPs into DM from ALP-strahlung production for ALP masses up to $m_{a}=8.5$\,GeV. Signal Monte Carlo events have been generated using \hbox{\texttt{MadGraph5\,v2.2.2}} \cite{Alwall:2014hca}. We have generated samples using a fixed ALP mass per sample in steps of 0.05\,GeV with 10,000 events each, using a branching ratio into DM of $\text{BR}(a\to\chi \chi)=1.0$. The final state consists of a single, highly energetic photon with an energy
\begin{equation}
\label{eq:recoilenergy}
E_{\gamma} = \frac{s-m_{a}^2}{2\sqrt{s}} \; ,
\end{equation}
where $\sqrt{s}=10.58$\,GeV is the collision energy. This search is very similar to the search of Dark Photon decays into DM described in ref.~\cite{lit:b2tip}. The backgrounds for this search have been found to be due to high cross section QED processes $e^+e^-\to e^+e^-\gamma(\gamma)$ and $e^+e^-\to\gamma\gamma(\gamma)$ where all but one photon are undetected. The background composition is a complicated function of detector geometry details that cannot be adequately reproduced without a full \belletwo detector simulation. We therefore take the background rates from ref.~\cite{lit:b2tip}. It should be noted that the irreducible background from $e^+e^-\to \nu \bar{\nu}\gamma$ is negligible. We obtain the signal efficiency for ALPs using generator-level Monte Carlo simulations.

\begin{figure}
\centering
\includegraphics[width=0.45\textwidth]{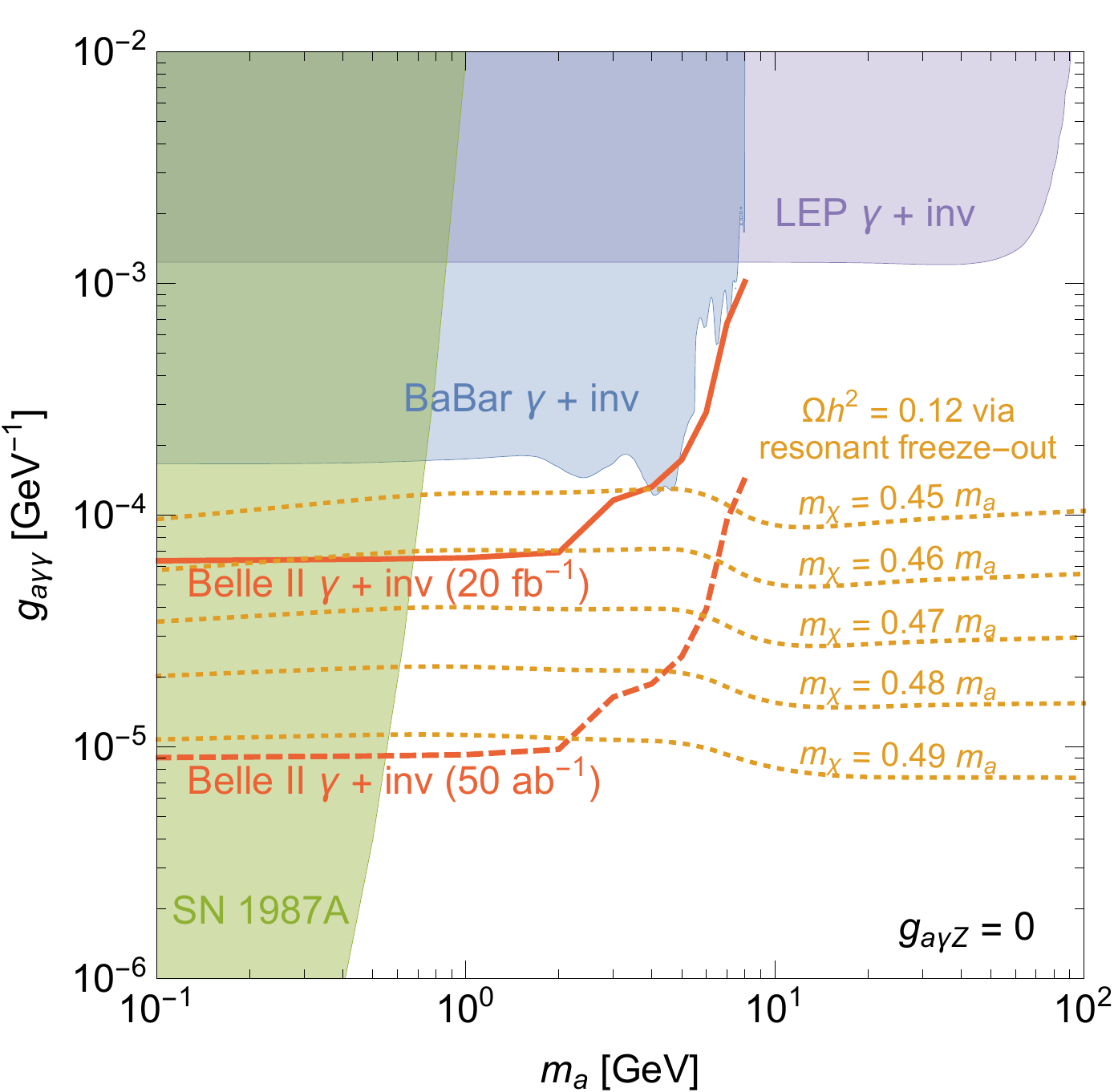}
\caption{\label{fig:BelleII_inv} Present and future constraints on ALPs decaying into DM compared to the parameter region where one can reproduce the observed DM relic abundance via resonant annihilation of DM into photons. Note that this process is efficient only if $m_\chi$ is slightly smaller than $m_a / 2$ (see figure~\ref{fig:sigmav}).}
\end{figure}

We determine the expected 90\,\% CL upper limit of signal events $n_s$ such that the Poisson probability of observing less than $n$ events when expecting $n_s+n_b$ events is $\leq0.1$, where $n$ is the integer closest to the number of background events $n_b$. Expected upper limits on the coupling $\ga$ are summarized as a function of ALP mass $m_{a}$ in figure~\ref{fig:BelleII_inv}. The much better expected sensitivity compared to \babar is mainly due to the more homogeneous calorimeter of \belletwo. Figure~\ref{fig:BelleII_inv} also shows the parameter ranges corresponding to resonant freeze-out. We observe that, if DM annihilation into photons is resonantly enhanced, existing experiments are not yet sensitive to the values of $\ga$ implied by the observed DM relic abundance, but \belletwo has a unique potential to probe the parameter regions of particular interest.

The sensitivity to high mass ALPs is limited by the trigger threshold for a single photon that will be implemented in \belletwo. We conservatively assume a trigger energy threshold of 1.8\,GeV which limits the search to ALP masses below $m_{a}$=8.6\,GeV. If the trigger threshold can be lowered to $1.2\,\text{GeV}$, the sensitivity extends to ALP masses up to $m_{a}$=9.3\,GeV. A higher collision energy close to the $\Upsilon(6S)$ resonance could further extend the sensitivity to about  $m_{a}$=9.7\,GeV for a trigger threshold of 1.2\,GeV. 

It should be noted that while the dominant physics background for this study comes from $e^+e^-\to \gamma\gamma(\gamma)$ events, the largest fraction of the trigger rate for trigger thresholds $\lesssim 1.8$\,GeV is due to radiative Bhabha events $e^+e^-\to e^+e^-\gamma(\gamma)$  where both tracks are out of the detector acceptance. 

\subsection{ALP decays into two photons}

\begin{figure}
\centering
\includegraphics[width=0.45\textwidth]{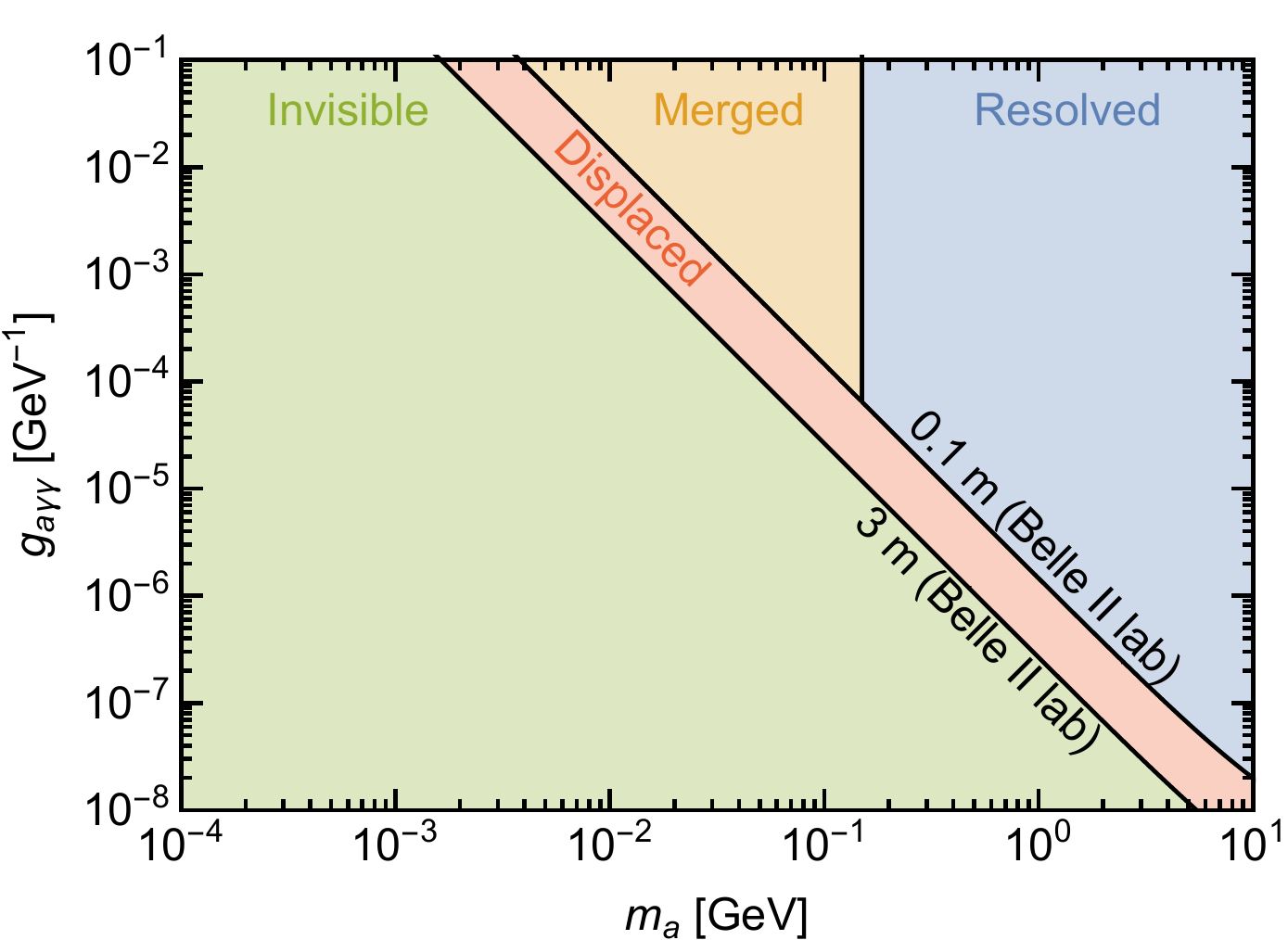}
\caption{\label{fig:regimes} Illustration of the different kinematic regimes relevant for ALP decays into two photons with \belletwo.}
\end{figure}

The experimental signature of the decays into two photons is determined by the relation between mass and coupling of the ALP. This relation affects both the decay length of the ALP and the opening angle of the decay photons. It leads to four different experimental signatures (see figure~\ref{fig:regimes}):
\begin{enumerate}
\item  ALPs with a mass of $\mathcal{O(\text{GeV})}$ decay promptly, and the opening angle of the decay photons is large enough that both decay photons can be resolved in the \belletwo detector (\emph{resolved}).
\item For lighter ALP masses but large couplings $\ga$, the decay is prompt but the ALP is highly boosted and the decay photons merge into one reconstructed cluster in  \belletwo calorimeter if $m_a \lesssim 150$\,MeV (\emph{merged}).\footnote{This corresponds to an average opening angle of about $(3-5)^{\circ}$ in the lab system that depends on the position in the detector.}
\item Even lighter ALPs  decay displaced from the interaction point but still inside the \belletwo detector. This is a challenging signature that consists of two reconstructed clusters, one of which has a \emph{displaced} vertex and contains two merged photons. The latter two conditions typically yield a bad quality of the reconstructed photon candidate which is not included in \textit{resolved} searches with final state photons. There is however enough detector activity in the ECL or KLM that these are vetoed in searches for \textit{invisible} final states to reduce high rate $e^+e^-\rightarrow\gamma\gamma$ backgrounds.
\item The lifetime of light ALPs with small couplings is large enough that a significant fraction of ALPs decays outside of \belletwo. The experimental signature is a single photon final state that looks identical to the ALP decay into DM (\emph{invisible}).
\end{enumerate}
The detailed \belletwo sensitivity for \emph{merged} and \emph{displaced} decays depends strongly on the actual performance of the \belletwo reconstruction software and beam background levels which are beyond the scope of this paper. It may be possible to search for displaced clusters in the KLM due to its longitudinal resolution of $\mathcal{O}$(10\,cm). The \textit{resolved} region can potentially be extended towards smaller ALP masses if the ECL reconstruction is improved.

We study \emph{resolved} ALP decays into two photons ($e^+e^-\to \gamma (a\to\gamma\gamma)$) over a range of ALP masses between $m_{a}$\,=\,(0.05--9.0)\,GeV. Signal Monte Carlo events have been generated using \hbox{\texttt{MadGraph5\,v2.2.2}}~\cite{Alwall:2014hca}. We have generated samples using a fixed ALP mass per sample in steps of 0.05\,GeV with 10,000 events each using a branching ratio into photons of $\text{BR}(a\to\gamma\gamma)=1.0$.

We use a simplified geometry description of the \belletwo detector to take into account the ECL geometry acceptance. We use the ECL energy resolution \cite{lit:b2tip} that is expected for 10\,\% of the full luminosity backgrounds. It should be noted that all three photons have rather high energy and the photon energy resolution is expected to not change significantly for even higher beam backgrounds. We use the approximate ECL crystal positions and sizes to estimate the performance in resolving overlapping photon clusters.

We find that the background is dominated by the QED process $e^+e^-\to\gamma\gamma\gamma$ with three photons in the final state. Background samples are generated using BABAYAGA.NLO \cite{CarloniCalame:2003yt, CarloniCalame:2001ny, CarloniCalame:2000pz}. Additional small backgrounds for small ALP masses may arise from $e^+e^-\to\gamma\gamma$ with a third photon candidate coming from beam-induced backgrounds, and from $e^+e^-\to\gamma\gamma$ where one of the photon converts into an electron-positron pair outside of the tracking detectors. The former will be reduced using the very good time resolution $\mathcal{O}$(ns) of the \belletwo ECL at high photon energies \cite{Aulchenko:2017lmh}. To reduce background from pair conversion, one can use the fact that the secondary electron-positron pair splits in the magnetic field and veto events where the polar angle difference between the photons of the lowest invariant mass photon pair is small and the azimuthal angle difference is rather large. We expect that the pair conversion background is only relevant in the ECL backward region where significant material from the CDC readout electronics is placed about 40\,cm away from the crystal front. 

A further potential background arises from the SM processes $e^+e^- \to \pi^0 \gamma$, $e^+e^- \to \eta \gamma$ and  $e^+e^- \to \eta' \gamma$~\cite{Czyz:2017veo}. We therefore exclude $_{-75\,\text{MeV}}^{+50\,\text{MeV}}$ mass regions around the nominal $\eta$ and $\eta'$ masses (our analysis is not sensitive to $m_a \approx m_\pi$). In the actual analysis a full study of these backgrounds should be included. Finally, we assume that both beam backgrounds and pair conversion backgrounds can be reduced to a negligible level using the selections described above, without significantly affecting the signal selection efficiency. 

Our event selection requires three photons with a CM energy $E^* > 0.25$\,GeV and a polar angle in the laboratory frame $17^{\circ} < \theta^{\text{lab}} < 150^{\circ}$. The invariant mass of the two photons from the ALP decay will peak at the ALP mass. We perform the sensitivity study twice, once using all three possible photon pair combinations (\textit{high mass selection}) and once using only the photon pair combination with the lowest invariant mass (\textit{low mass selection}). The latter has a smaller signal efficiency especially at higher ALP masses but a lower combinatorial background. For the three photon combination case we select events where the maximum absolute cosine of the three helicity angles is less than 0.9, and for the two photon combination case we keep events where the absolute cosine of the helicity angle is less than 0.6. These selection criteria maximize the ratio of $\sqrt{S}/B$, where $S$ is the number of signal events and $B$ is the number of background events, after all other selection criteria have been applied. It should be noted that the helicity selection criteria not only reduce $e^+e^-\to\gamma\gamma\gamma$ backgrounds, but will also suppress backgrounds from $e^+e^-\to\gamma\gamma$ combined with a random third photon from beam backgrounds. We require that all three photons are separated by at least 2\,ECL crystals in both polar and azimuthal direction. We do not constrain the three photon invariant mass to the collision energy since our MadGraph signal Monte Carlo does not include additional photon radiation whereas the background Monte Carlo does.

\begin{figure}
\centering
\includegraphics[width=0.45\textwidth]{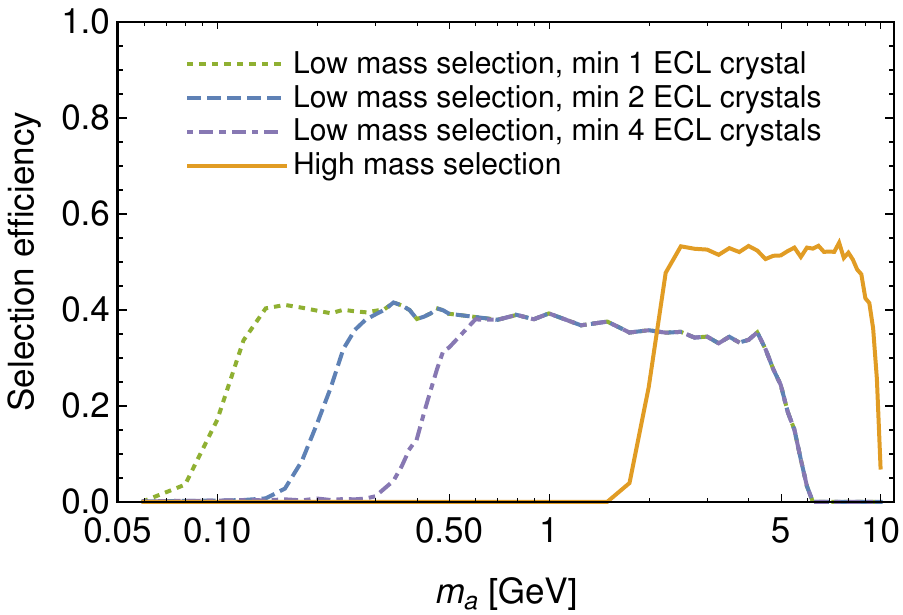}
\caption{\label{fig:belle2_eff} \belletwo $3\gamma$ efficiency as function of ALP mass after the final selection. The different \textit{low mass} selections correspond to a minimum photon separation of 1, 2, and 4 crystals in the ECL which is an approximation for the expected performance of an improved reconstruction, the default reconstruction and the reconstruction in the first trigger level (see text for details).}
\end{figure}

We finally select candidates within $[-3\sigma_{m_2}, +1.5\sigma_{m_2}]$ around the generated ALP mass, where $\sigma_{m_2}$ is the invariant mass resolution of the decay photon pair. For high mass ALPs we select events within $[-3\sigma_{\gamma}, +1.5\sigma_{\gamma}]$ around the expected recoil photon energy (see equation \ref{eq:recoilenergy}) instead. The ranges contains about 85\,\% of the previously selected signal events. The signal efficiency after all selections is flat and about (35--40)\,\% ((50--55)\,\%) for the two photon (three photon) combination (see figure~\ref{fig:belle2_eff}). The photon angle separation distance of 2\,ECL crystals is a conservative estimate of the \belletwo offline reconstruction performance and can likely be improved using advanced reconstruction techniques based on Machine Learning methods, and by using shower shape techniques similar to those applied in high energy $\pi^0$ reconstruction. We show the efficiency for single ECL crystal difference for comparison as well.

Events from $e^+e^-\to \gamma (a\to\gamma\gamma)$ are typically triggered by three energy depositions of at least 0.1\,GeV in the ECL. Unlike in the \belletwo offline reconstruction, the photon reconstruction at trigger level is much simpler and has a worse angular separation power. We expect that a separation of less than 4 ECL crystals will result in merged photon clusters and make this trigger inefficient for ALP masses below about 0.5\,GeV. An ideal trigger will require at least two highly energetic ECL clusters and must not satisfy  $e^+e^-\to e^+e^-$  (Bhabha) vetoes. However, any $e^+e^-\to\gamma\gamma$ veto decision must be delayed to the high level trigger where offline reconstruction is available in order to maintain a high trigger efficiency for low mass ALPs.

\begin{figure}
\centering
\includegraphics[width=0.45\textwidth]{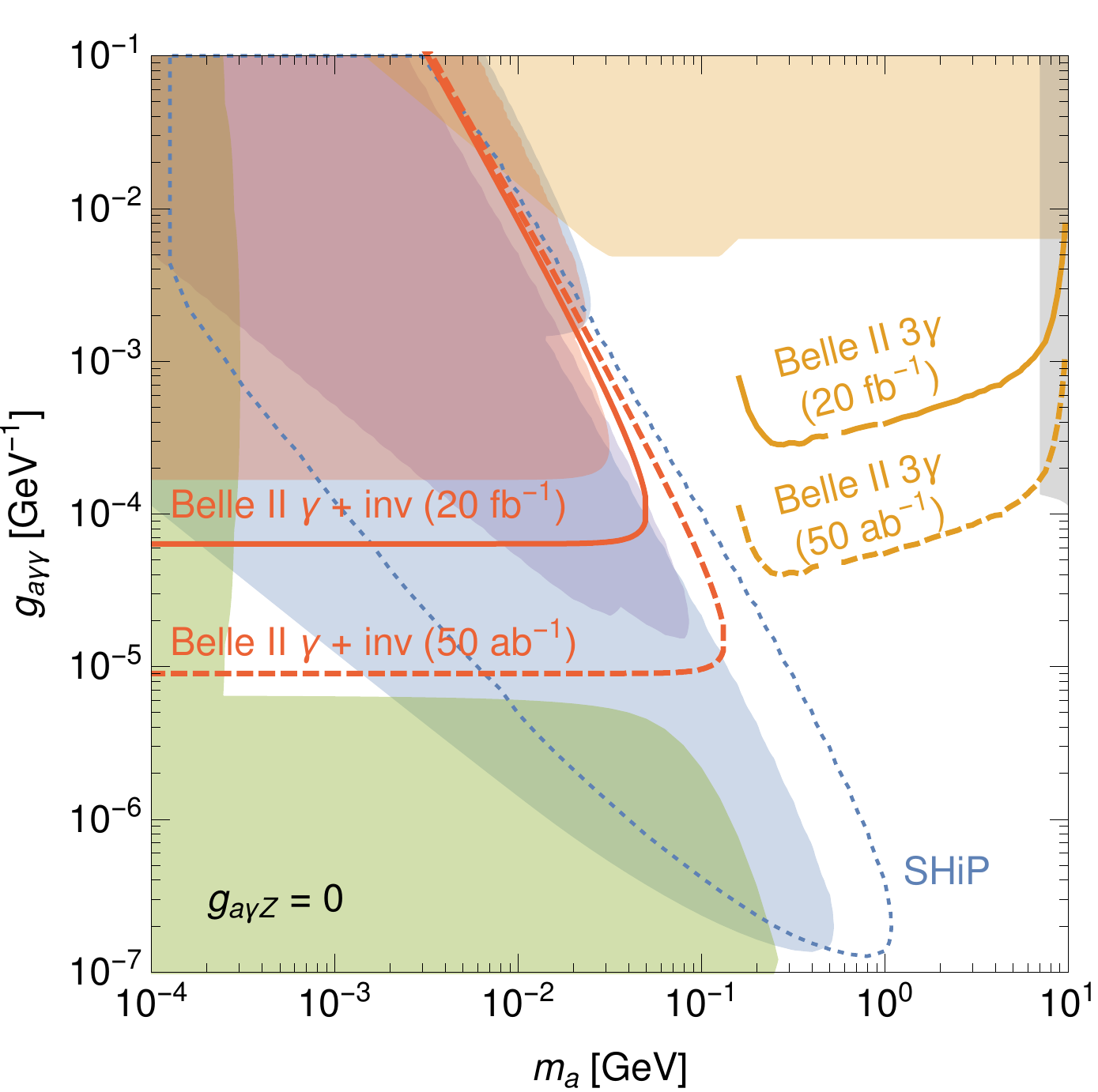}\qquad
\includegraphics[width=0.45\textwidth]{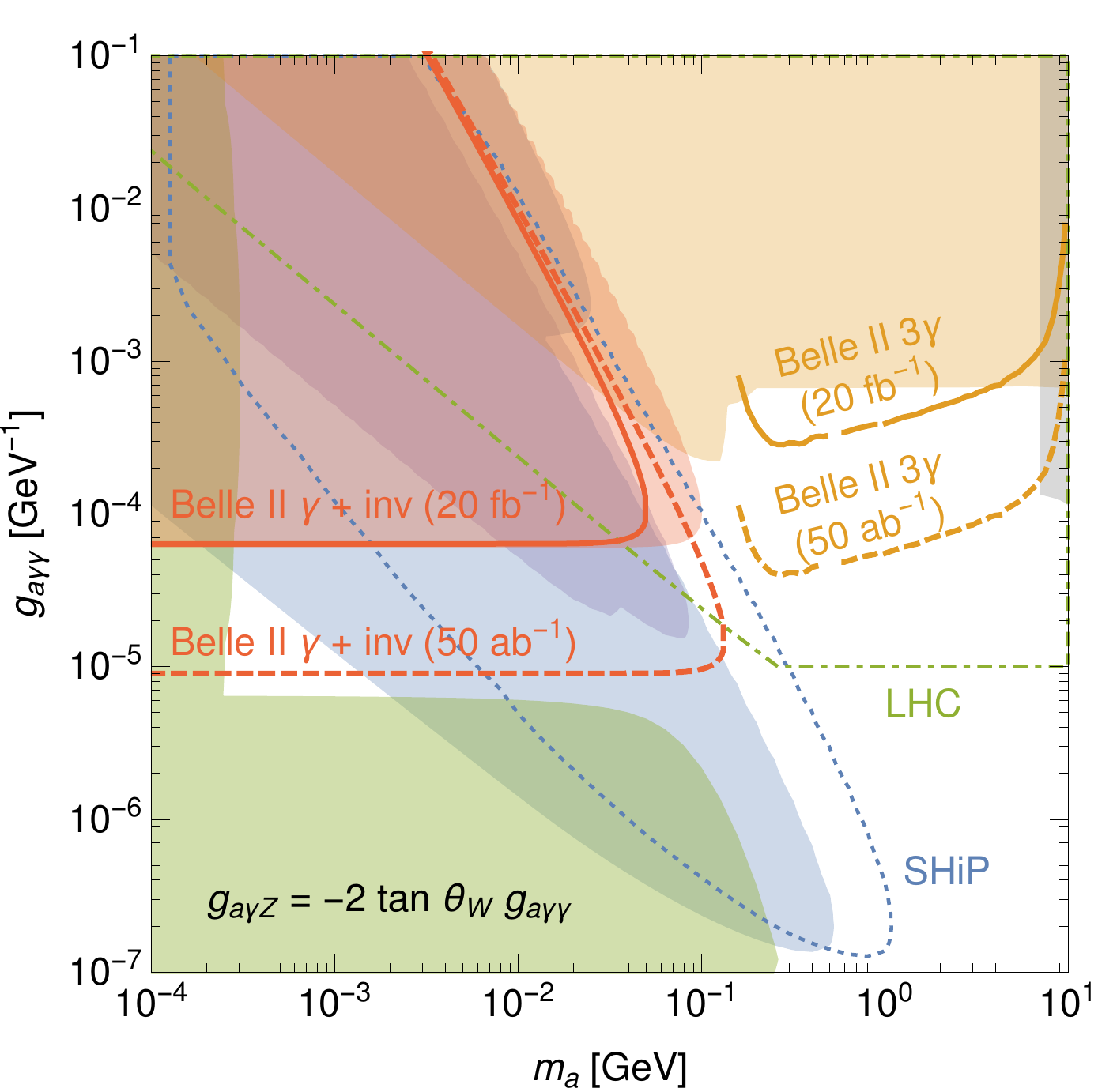}
\caption{\label{fig:BelleII} Projected \belletwo sensitivity (90\,\% CL) compared to existing constraints on ALPs with photon coupling (left) and hypercharge coupling (right), as well as the projected sensitivities from SHiP~\cite{Dobrich:2015jyk} and the LHC~\cite{Bauer:2017ris}.}
\end{figure}

We obtain the expected 90\,\% CL sensitivity as described above. The sensitivity for long-lived ALPs decaying into two photons is determined from the sensitivity of ALP decays into DM, taking into account the reduced efficiency given by eq.~(\ref{eq:decay}) using a detector length\footnote{The event selection includes a veto of energy depositions in the KLM. The detector length is hence taken as approximate outer radius of the barrel KLM.} of $L_\mathrm{D}=300$\,cm \cite{Abashian:2000cg}. The projected sensitivities to the coupling $\ga$ are summarized as a function of ALP mass $m_{a}$ in figure~\ref{fig:BelleII}.

We make a number of important observations from figure~\ref{fig:BelleII}. First of all, we note that for very light ALPs (i.e. $m_a \sim 1\,\mathrm{MeV}$) \belletwo single-photon searches can push significantly beyond current constraints from beam dump experiments and can potentially explore the triangular region around $\ga \sim 10^{-5}\,\mathrm{GeV^{-1}}$, which is currently only constrained by model-dependent cosmological considerations. For heavier ALPs (i.e. $150\,\mathrm{MeV} < m_a < 10\,\mathrm{GeV}$) \belletwo searches for three resolved photons can significantly improve over existing bounds from LEP even with early data ($20\,\mathrm{fb^{-1}}$). With larger data sets, \belletwo will be able to probe couplings of the order of $\ga \sim 10^{-4}\,\mathrm{GeV^{-1}}$ over a wide range of ALP masses. Improvements in the \belletwo reconstruction software could push the sensitivity for three resolved photons to slightly lower masses and a dedicated search for displaced photons could extend the long-lived search towards higher masses.

Comparing the two panels in figure~\ref{fig:BelleII} we note furthermore that there is a remarkable complementarity between \belletwo, SHiP and LHC. SHiP will have greatest sensitivity in the parameter region where the ALP decay length is $\mathcal{O}(1\text{--}100)\,\mathrm{m}$, which is difficult to explore with \belletwo and the LHC. The LHC, on the other hand, is sensitive mostly to the coupling $g_{a\gamma Z}$, while \belletwo and SHiP directly probe the ALP-photon coupling $\ga$. The combination of these experiments will therefore allow to make significant progress in the exploration of the ALP parameter space. Moreover, we can hope to see an ALP signal in more than one experiment, which would potentially enable us to reconstruct its properties and coupling structure.

\subsection{Photon fusion}
\label{sec:photonfusion}

So far we have focused on the case that the ALP is produced in association with a highly-energetic photon, which facilitates an efficient reconstruction of these events. For ALPs produced in photon fusion the situation becomes more complicated, as the transverse momenta of electron and positron after the collision (and hence their polar angle) are too small to be detectable.

\begin{figure}
\centering
\includegraphics[width=0.4\textwidth]{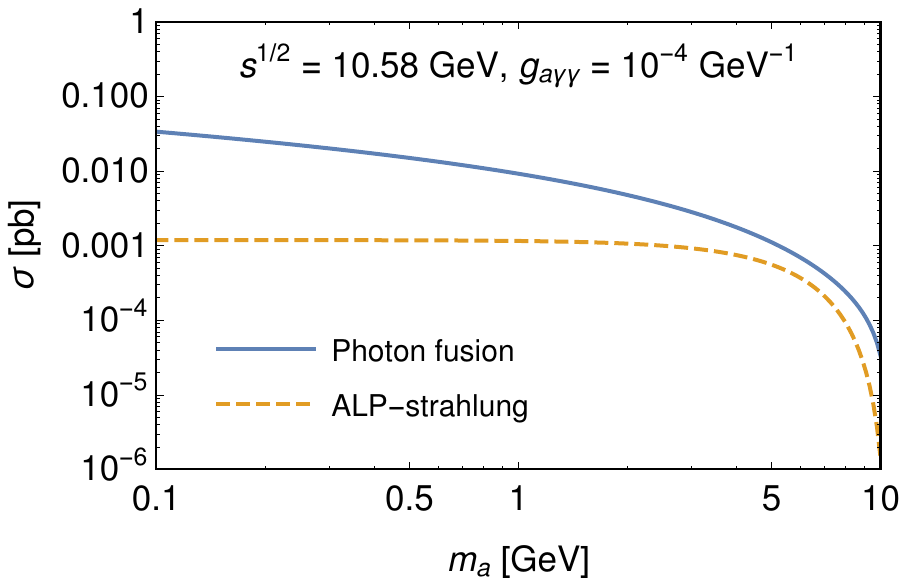}\qquad
\includegraphics[width=0.4\textwidth]{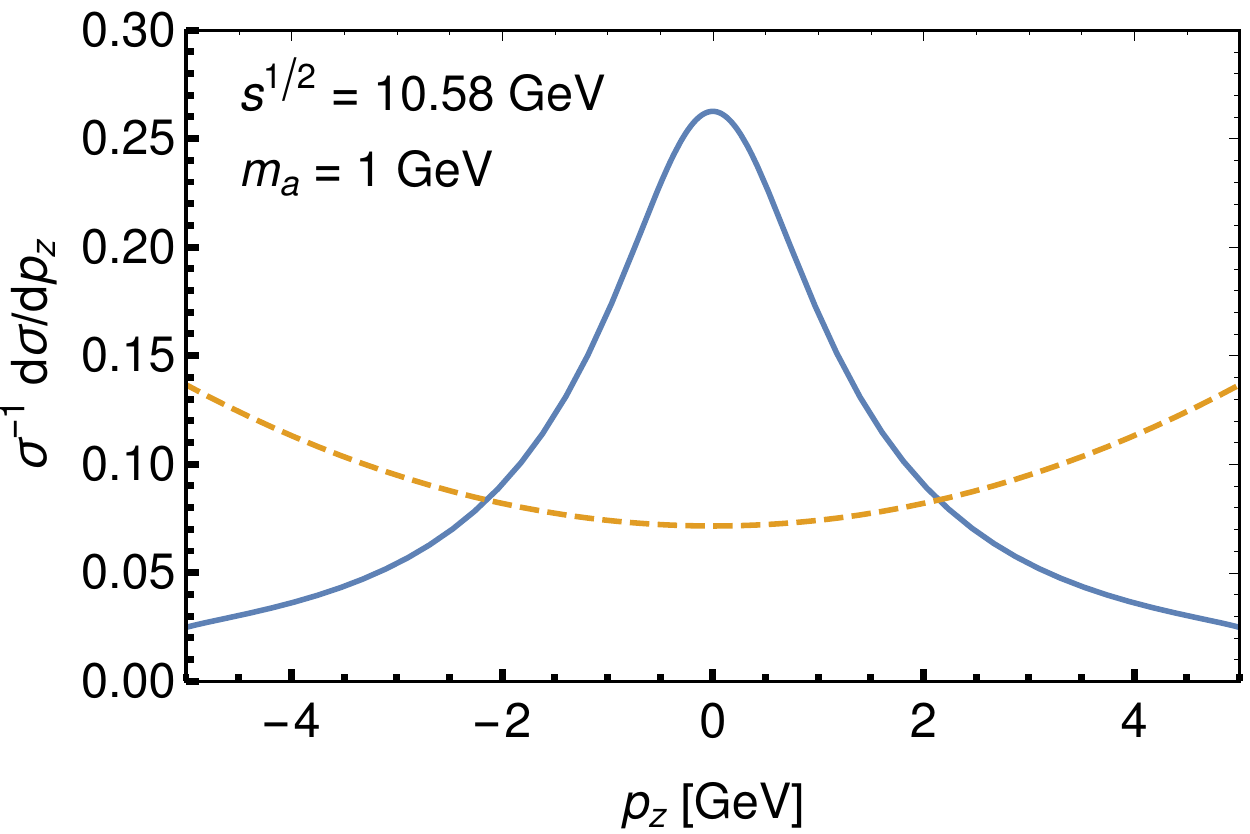}
\caption{\label{fig:fusion} Comparison of ALP production in $e^+ e^-$ collisions via ALP-strahlung and via photon fusion. The left panel shows the total cross section, the right panel the differential cross section with respect to the longitudinal momentum $p_z$.}
\end{figure}

Searches for ALPs produced in photon fusion are interesting for two reasons: First, as shown in the left panel of figure~\ref{fig:fusion} the total ALP production cross section from photon fusion significantly exceeds the one from ALP-strahlung (in particular for small ALP masses), so that photon fusion is responsible for the vast majority of ALPs produced at \belletwo~\cite{Marciano:2016yhf,Izaguirre:2016dfi}. And second, the production cross section from photon fusion peaks for small ALP momenta, i.e.\ ALPs will be produced dominantly at rest (see right panel of figure~\ref{fig:fusion}). This means that, in contrast to ALP-strahlung, the opening angle between the two photons produced in the ALP decay will typically be large even for low-mass ALPs. 

The signature in the \belletwo detector will consist of two photons with an invariant mass equal to the ALP mass and missing energy along the beam-pipe. The azimuthal angles of the two photons are back-to-back in the CM frame. The \belletwo acceptance for ALPs produced in photon fusion is high: For $m_a=0.2$\,GeV ($m_a=2.0$\,GeV) 66\,\% (89\,\%) of all ALPs have both decay photons in the ECL acceptance. However, for low mass ALPs the photon energy is small and often below a typical trigger threshold of 100\,MeV per ECL cluster. Studies have shown a very large beam-induced background of low energy ECL clusters \cite{lit:b2tip}, making the detection of ALPs produced in photon fusion very difficult. While it may be possible to reject a part of these events in the offline reconstruction, the trigger rates are probably too large to sustain. 

A possible opportunity might be to consider the case where either the electron or the positron receives sufficiently large transverse momentum to be tagged~\cite{Izaguirre:2016dfi}. However, the cross section is strongly reduced in this case and backgrounds remain rather large compared to the $3\gamma$ final state. A detailed study of the potential sensitivity of \belletwo in this search channel is beyond the scope of this work. 

\section{Conclusions}
\label{sec:conclusions}

In this work we have discussed the experimental situation concerning axion-like particles (ALPs) that interact with Standard Model particles dominantly via couplings to photons and electroweak gauge bosons. Reviewing existing constraints we have argued that the bounds from $e^+$-$e^-$ colliders conventionally shown for this scenario are outdated. We have updated these constraints by reinterpreting mono-photon searches at LEP and Dark Photon searches at \babar. We have furthermore investigated the bounds on ALPs from electron beam dumps and from SN 1987A and provided refined estimates of these constraints. We have also discussed the case that ALPs can be produced in the decay $Z \to a \gamma$, which is relevant if ALPs couple to hypercharge. A summary of existing constraints is shown in figure~\ref{fig:constraints}.

A scenario of particular interest is ALPs coupled to a light DM particle, which induces invisible ALP decays and hence removes a number of experimental constraints. In this case DM can pair-annihilate into photons via ALP exchange. We pointed out that if these annihilations are resonantly enhanced in the Early Universe, the DM particle can be a thermal relic with the required abundance and satisfy all experimental and observational constraints. The DM relic abundance then depends on only two parameters (the ALP-photon coupling and the ratio of DM mass to ALP mass), leading to a highly predictive scenario. Depending on how close the mass ratio is to resonance ($m_\chi \sim m_a / 2$), the observed DM relic abundance implies ALP-photon couplings in the range $\ga \sim 10^{-4}\text{--}10^{-5}\,\mathrm{GeV}^{-1}$. These values are consistent with all existing experimental constraints but lie precisely in the parameter region that can be probed with single-photon searches at \belletwo.

Our central observation is that existing bounds on ALPs can be significantly improved with single-photon searches and searches for three resolved photons at \belletwo. The former type of search probes invisibly decaying ALPs as well as ALPs that escape from the detector before decaying, while the latter search strategy is highly sensitive to ALPs in the GeV region, which decay promptly into a pair of photons. We have shown that the sensitivity of single-photon searches is substantially better than has been estimated previously due to a significant reduction in background compared to \babar. In combination these search strategies can cover wide ranges of ALP parameter space and explore various models that are of interest for both cosmology and particle physics (see figures~\ref{fig:BelleII_inv} and~\ref{fig:BelleII}). We have found searches with \belletwo to be highly complementary to searches with SHiP (which probes different ALP decay lengths) and at the LHC (which probes different coupling structures).

We have pointed out that there are two regimes that require additional studies with full \belletwo simulations in order to understand the sensitivity: ALP decays from displaced vertices and prompt decays of highly-boosted ALPs leading to a merging of the two photons. Finding ways to extend the \belletwo sensitivity in these regions will be of great interest for future work. Searching for ALPs produced via photon fusion is one potential avenue to make progress, but backgrounds for this search channel may be prohibitive. Future progress in the search for ALPs therefore requires both the implementation of the search strategies described in this work as well as the exploration of innovative approaches that can reach the remaining corners of parameter space. Combining these with a better understanding of astrophysical constraints of ALPs may finally enable us to understand the role that ALPs play in the Universe.

\acknowledgments

We would like to thank Babette D\"obrich and Joerg Jaeckel for helpful comments on the manuscript and Rebecca Leane, Josef Pradler, Georg Raffelt, Diego Redigolo, Javier Redondo, Brian Shuve and Martin W.~Winkler for useful discussions. This work is supported by the German Science Foundation (DFG) under the Collaborative Research Center (SFB) 676 ``Particles, Strings and the Early Universe'' and the Emmy Noether Grant No.\ KA 4662/1-1, the ERC Starting Grant `NewAve' (638528) as well as NSERC (Canada).

\providecommand{\href}[2]{#2}\begingroup\raggedright\endgroup


\begin{thebibliography}{10}

\bibitem{Hook:2014cda}
A.~Hook, \href{http://dx.doi.org/10.1103/PhysRevLett.114.141801}{{\it
  {Anomalous solutions to the strong CP problem}}, } {\em Phys. Rev. Lett.}
  {\bf 114} (2015), no.~14 141801, [\href{http://arxiv.org/abs/1411.3325}{{\tt
  1411.3325}}].

\bibitem{Fukuda:2015ana}
H.~Fukuda, K.~Harigaya, M.~Ibe, and T.~T. Yanagida,
  \href{http://dx.doi.org/10.1103/PhysRevD.92.015021}{{\it {Model of visible
  QCD axion}}, } {\em Phys. Rev.} {\bf D92} (2015), no.~1 015021,
  [\href{http://arxiv.org/abs/1504.06084}{{\tt 1504.06084}}].

\bibitem{Arvanitaki:2009fg}
A.~Arvanitaki, S.~Dimopoulos, S.~Dubovsky, N.~Kaloper, and J.~March-Russell,
  \href{http://dx.doi.org/10.1103/PhysRevD.81.123530}{{\it {String Axiverse}},
  } {\em Phys. Rev.} {\bf D81} (2010) 123530,
  [\href{http://arxiv.org/abs/0905.4720}{{\tt 0905.4720}}].

\bibitem{Cicoli:2012sz}
M.~Cicoli, M.~Goodsell, and A.~Ringwald,
  \href{http://dx.doi.org/10.1007/JHEP10(2012)146}{{\it {The type IIB string
  axiverse and its low-energy phenomenology}}, } {\em JHEP} {\bf 10} (2012)
  146, [\href{http://arxiv.org/abs/1206.0819}{{\tt 1206.0819}}].
  
\bibitem{Bellazzini:2017neg}
  B.~Bellazzini, A.~Mariotti, D.~Redigolo, F.~Sala and J.~Serra,
  {\it {R-axion at colliders}}, \href{http://arxiv.org/abs/1702.02152}{{\tt 1702.02152}}.

\bibitem{Hewett:2012ns}
J.~L. Hewett et~al.,
  \href{http://inspirehep.net/record/1114323/files/arXiv:1205.2671.pdf}{{\it
  {Fundamental Physics at the Intensity Frontier}}, } 2012.
\newblock \href{http://arxiv.org/abs/1205.2671}{{\tt 1205.2671}}.

\bibitem{Essig:2013lka}
R.~Essig et~al.,
  \href{http://inspirehep.net/record/1263039/files/arXiv:1311.0029.pdf}{{\it
  {Working Group Report: New Light Weakly Coupled Particles}}, } 2013.
\newblock \href{http://arxiv.org/abs/1311.0029}{{\tt 1311.0029}}.

\bibitem{Abe:2010gxa}
{\bf Belle\,II Collaboration}, T.~Abe et~al., {\it {Belle II Technical Design
  Report}},  \href{http://arxiv.org/abs/1011.0352}{{\tt 1011.0352}}.

\bibitem{NA62:2017rwk}
{\bf NA62 Collaboration}, E.~Cortina~Gil et~al.,
  \href{http://dx.doi.org/10.1088/1748-0221/12/05/P05025}{{\it {The Beam and
  detector of the NA62 experiment at CERN}}, } {\em JINST} {\bf 12} (2017),
  no.~05 P05025, [\href{http://arxiv.org/abs/1703.08501}{{\tt 1703.08501}}].

\bibitem{Anelli:2015pba}
{\bf SHiP Collaboration}, M.~Anelli et~al., {\it {A facility to Search for
  Hidden Particles (SHiP) at the CERN SPS}},
  \href{http://arxiv.org/abs/1504.04956}{{\tt 1504.04956}}.

\bibitem{Cadamuro:2011fd}
D.~Cadamuro and J.~Redondo,
  \href{http://dx.doi.org/10.1088/1475-7516/2012/02/032}{{\it {Cosmological
  bounds on pseudo Nambu-Goldstone bosons}}, } {\em JCAP} {\bf 1202} (2012)
  032, [\href{http://arxiv.org/abs/1110.2895}{{\tt 1110.2895}}].

\bibitem{Millea:2015qra}
M.~Millea, L.~Knox, and B.~Fields,
  \href{http://dx.doi.org/10.1103/PhysRevD.92.023010}{{\it {New Bounds for
  Axions and Axion-Like Particles with keV-GeV Masses}}, } {\em Phys. Rev.}
  {\bf D92} (2015), no.~2 023010, [\href{http://arxiv.org/abs/1501.04097}{{\tt
  1501.04097}}].

\bibitem{Arias:2012az}
P.~Arias, D.~Cadamuro, M.~Goodsell, J.~Jaeckel, J.~Redondo, et~al.,
  \href{http://dx.doi.org/10.1088/1475-7516/2012/06/013}{{\it {WISPy Cold Dark
  Matter}}, } {\em JCAP} {\bf 1206} (2012) 013,
  [\href{http://arxiv.org/abs/1201.5902}{{\tt 1201.5902}}].

\bibitem{Giannotti:2015kwo}
M.~Giannotti, I.~Irastorza, J.~Redondo, and A.~Ringwald,
  \href{http://dx.doi.org/10.1088/1475-7516/2016/05/057}{{\it {Cool WISPs for
  stellar cooling excesses}}, } {\em JCAP} {\bf 1605} (2016), no.~05 057,
  [\href{http://arxiv.org/abs/1512.08108}{{\tt 1512.08108}}].

\bibitem{Meyer:2013pny}
M.~Meyer, D.~Horns, and M.~Raue,
  \href{http://dx.doi.org/10.1103/PhysRevD.87.035027}{{\it {First lower limits
  on the photon-axion-like particle coupling from very high energy gamma-ray
  observations}}, } {\em Phys. Rev.} {\bf D87} (2013), no.~3 035027,
  [\href{http://arxiv.org/abs/1302.1208}{{\tt 1302.1208}}].

\bibitem{Cicoli:2014bfa}
M.~Cicoli, J.~P. Conlon, M.~C.~D. Marsh, and M.~Rummel,
  \href{http://dx.doi.org/10.1103/PhysRevD.90.023540}{{\it {3.55\,keV photon
  line and its morphology from a 3.55\,keV axionlike particle line}}, } {\em
  Phys. Rev.} {\bf D90} (2014) 023540,
  [\href{http://arxiv.org/abs/1403.2370}{{\tt 1403.2370}}].

\bibitem{Conlon:2014xsa}
J.~P. Conlon and F.~V. Day,
  \href{http://dx.doi.org/10.1088/1475-7516/2014/11/033}{{\it {3.55\,keV photon
  lines from axion to photon conversion in the Milky\,Way and M31}}, } {\em
  JCAP} {\bf 1411} (2014) 033, [\href{http://arxiv.org/abs/1404.7741}{{\tt
  1404.7741}}].

\bibitem{Jaeckel:2014qea}
J.~Jaeckel, J.~Redondo, and A.~Ringwald,
  \href{http://dx.doi.org/10.1103/PhysRevD.89.103511}{{\it {3.55\,keV hint for
  decaying axionlike particle dark matter}}, } {\em Phys. Rev.} {\bf D89}
  (2014) 103511, [\href{http://arxiv.org/abs/1402.7335}{{\tt 1402.7335}}].

\bibitem{Mimasu:2014nea}
K.~Mimasu and V.~Sanz, \href{http://dx.doi.org/10.1007/JHEP06(2015)173}{{\it
  {ALPs at Colliders}}, } {\em JHEP} {\bf 06} (2015) 173,
  [\href{http://arxiv.org/abs/1409.4792}{{\tt 1409.4792}}].

\bibitem{Dolan:2014ska}
M.~J. Dolan, F.~Kahlhoefer, C.~McCabe, and K.~Schmidt-Hoberg,
  \href{http://dx.doi.org/10.1007/JHEP07(2015)103,
  10.1007/JHEP03(2015)171}{{\it {A taste of dark matter: Flavour constraints on
  pseudoscalar mediators}}, } {\em JHEP} {\bf 03} (2015) 171,
  [\href{http://arxiv.org/abs/1412.5174}{{\tt 1412.5174}}]. [Erratum:
  JHEP07,103(2015)].

\bibitem{Jaeckel:2015jla}
J.~Jaeckel and M.~Spannowsky,
  \href{http://dx.doi.org/10.1016/j.physletb.2015.12.037}{{\it {Probing MeV to
  90\,GeV axion-like particles with LEP and LHC}}, } {\em Phys. Lett.} {\bf
  B753} (2016) 482--487, [\href{http://arxiv.org/abs/1509.00476}{{\tt
  1509.00476}}].

\bibitem{Dobrich:2015jyk}
B.~D{\"o}brich, J.~Jaeckel, F.~Kahlhoefer, A.~Ringwald, and K.~Schmidt-Hoberg,
  \href{http://dx.doi.org/10.1007/JHEP02(2016)018}{{\it {ALPtraum: ALP
  production in proton beam dump experiments}}, } {\em JHEP} {\bf 02} (2016)
  018, [\href{http://arxiv.org/abs/1512.03069}{{\tt 1512.03069}}].
  [JHEP02,018(2016)].

\bibitem{Izaguirre:2016dfi}
E.~Izaguirre, T.~Lin, and B.~Shuve,
  \href{http://dx.doi.org/10.1103/PhysRevLett.118.111802}{{\it {Searching for
  Axionlike Particles in Flavor-Changing Neutral Current Processes}}, } {\em
  Phys. Rev. Lett.} {\bf 118} (2017), no.~11 111802,
  [\href{http://arxiv.org/abs/1611.09355}{{\tt 1611.09355}}].

\bibitem{Knapen:2016moh}
S.~Knapen, T.~Lin, H.~K. Lou, and T.~Melia,
  \href{http://dx.doi.org/10.1103/PhysRevLett.118.171801}{{\it {Searching for
  Axion-like Particles with Ultraperipheral Heavy-Ion Collisions}}, } {\em
  Phys. Rev. Lett.} {\bf 118} (2017), no.~17 171801,
  [\href{http://arxiv.org/abs/1607.06083}{{\tt 1607.06083}}].

\bibitem{Brivio:2017ije}
I.~Brivio, M.~B. Gavela, L.~Merlo, K.~Mimasu, J.~M. No, et~al., {\it {ALPs
  Effective Field Theory and Collider Signatures}},
  \href{http://arxiv.org/abs/1701.05379}{{\tt 1701.05379}}.

\bibitem{Bauer:2017nlg}
M.~Bauer, M.~Neubert, and A.~Thamm,
  \href{http://dx.doi.org/10.1103/PhysRevLett.119.031802}{{\it {LHC as an Axion
  Factory: Probing an Axion Explanation for $(g-2)_\mu$ with Exotic Higgs
  Decays}}, } {\em Phys. Rev. Lett.} {\bf 119} (2017), no.~3 031802,
  [\href{http://arxiv.org/abs/1704.08207}{{\tt 1704.08207}}].

\bibitem{Choi:2017gpf}
K.~Choi, S.~H. Im, C.~B. Park, and S.~Yun, {\it {Minimal Flavor Violation with
  Axion-like Particles}},  \href{http://arxiv.org/abs/1708.00021}{{\tt
  1708.00021}}.

\bibitem{Bauer:2017ris}
M.~Bauer, M.~Neubert, and A.~Thamm, {\it {Collider Probes of Axion-Like
  Particles}},  \href{http://arxiv.org/abs/1708.00443}{{\tt 1708.00443}}.

\bibitem{Chang:2000ii}
D.~Chang, W.-F. Chang, C.-H. Chou, and W.-Y. Keung,
  \href{http://dx.doi.org/10.1103/PhysRevD.63.091301}{{\it {Large two loop
  contributions to g-2 from a generic pseudoscalar boson}}, } {\em Phys. Rev.}
  {\bf D63} (2001) 091301, [\href{http://arxiv.org/abs/hep-ph/0009292}{{\tt
  hep-ph/0009292}}].

\bibitem{Marciano:2016yhf}
W.~J. Marciano, A.~Masiero, P.~Paradisi, and M.~Passera,
  \href{http://dx.doi.org/10.1103/PhysRevD.94.115033}{{\it {Contributions of
  axionlike particles to lepton dipole moments}}, } {\em Phys. Rev.} {\bf D94}
  (2016), no.~11 115033, [\href{http://arxiv.org/abs/1607.01022}{{\tt
  1607.01022}}].

\bibitem{Ellwanger:2016wfe}
U.~Ellwanger and S.~Moretti,
  \href{http://dx.doi.org/10.1007/JHEP11(2016)039}{{\it {Possible Explanation
  of the Electron Positron Anomaly at 17\,MeV in $^8Be$ Transitions Through a
  Light Pseudoscalar}}, } {\em JHEP} {\bf 11} (2016) 039,
  [\href{http://arxiv.org/abs/1609.01669}{{\tt 1609.01669}}].

\bibitem{Flacke:2016szy}
T.~Flacke, C.~Frugiuele, E.~Fuchs, R.~S. Gupta, and G.~Perez,
  \href{http://dx.doi.org/10.1007/JHEP06(2017)050}{{\it {Phenomenology of
  Relaxion-Higgs mixing}}, } {\em JHEP} {\bf 06} (2017) 050,
  [\href{http://arxiv.org/abs/1610.02025}{{\tt 1610.02025}}].

\bibitem{Graham:2015cka}
P.~W. Graham, D.~E. Kaplan, and S.~Rajendran,
  \href{http://dx.doi.org/10.1103/PhysRevLett.115.221801}{{\it {Cosmological
  Relaxation of the Electroweak Scale}}, } {\em Phys. Rev. Lett.} {\bf 115}
  (2015), no.~22 221801, [\href{http://arxiv.org/abs/1504.07551}{{\tt
  1504.07551}}].

\bibitem{Nomura:2008ru}
Y.~Nomura and J.~Thaler,
  \href{http://dx.doi.org/10.1103/PhysRevD.79.075008}{{\it {Dark Matter through
  the Axion Portal}}, } {\em Phys. Rev.} {\bf D79} (2009) 075008,
  [\href{http://arxiv.org/abs/0810.5397}{{\tt 0810.5397}}].

\bibitem{Boehm:2014hva}
C.~Boehm, M.~J. Dolan, C.~McCabe, M.~Spannowsky, and C.~J. Wallace,
  \href{http://dx.doi.org/10.1088/1475-7516/2014/05/009}{{\it {Extended
  gamma-ray emission from Coy Dark Matter}}, } {\em JCAP} {\bf 1405} (2014)
  009, [\href{http://arxiv.org/abs/1401.6458}{{\tt 1401.6458}}].

\bibitem{Masso:1995tw}
E.~Masso and R.~Toldra, \href{http://dx.doi.org/10.1103/PhysRevD.52.1755}{{\it
  {On a light spinless particle coupled to photons}}, } {\em Phys. Rev.} {\bf
  D52} (1995) 1755--1763, [\href{http://arxiv.org/abs/hep-ph/9503293}{{\tt
  hep-ph/9503293}}].

\bibitem{Masso:1997ru}
E.~Masso and R.~Toldra, \href{http://dx.doi.org/10.1103/PhysRevD.55.7967}{{\it
  {New constraints on a light spinless particle coupled to photons}}, } {\em
  Phys. Rev.} {\bf D55} (1997) 7967--7969,
  [\href{http://arxiv.org/abs/hep-ph/9702275}{{\tt hep-ph/9702275}}].

\bibitem{Batell:2009di}
B.~Batell, M.~Pospelov, and A.~Ritz,
  \href{http://dx.doi.org/10.1103/PhysRevD.80.095024}{{\it {Exploring Portals
  to a Hidden Sector Through Fixed Targets}}, } {\em Phys. Rev.} {\bf D80}
  (2009) 095024, [\href{http://arxiv.org/abs/0906.5614}{{\tt 0906.5614}}].

\bibitem{Andreas:2012mt}
S.~Andreas, C.~Niebuhr, and A.~Ringwald,
  \href{http://dx.doi.org/10.1103/PhysRevD.86.095019}{{\it {New Limits on
  Hidden Photons from Past Electron Beam Dumps}}, } {\em Phys. Rev.} {\bf D86}
  (2012) 095019, [\href{http://arxiv.org/abs/1209.6083}{{\tt 1209.6083}}].

\bibitem{Essig:2013vha}
R.~Essig, J.~Mardon, M.~Papucci, T.~Volansky, and Y.-M. Zhong,
  \href{http://dx.doi.org/10.1007/JHEP11(2013)167}{{\it {Constraining Light
  Dark Matter with Low-Energy $e^+e^-$ Colliders}}, } {\em JHEP} {\bf 11}
  (2013) 167, [\href{http://arxiv.org/abs/1309.5084}{{\tt 1309.5084}}].

\bibitem{Izaguirre:2013uxa}
E.~Izaguirre, G.~Krnjaic, P.~Schuster, and N.~Toro,
  \href{http://dx.doi.org/10.1103/PhysRevD.88.114015}{{\it {New Electron
  Beam-Dump Experiments to Search for MeV to few-GeV Dark Matter}}, } {\em
  Phys. Rev.} {\bf D88} (2013) 114015,
  [\href{http://arxiv.org/abs/1307.6554}{{\tt 1307.6554}}].

\bibitem{Batell:2014mga}
B.~Batell, R.~Essig, and Z.~Surujon,
  \href{http://dx.doi.org/10.1103/PhysRevLett.113.171802}{{\it {Strong
  Constraints on Sub-GeV Dark Sectors from SLAC Beam Dump E137}}, } {\em Phys.
  Rev. Lett.} {\bf 113} (2014), no.~17 171802,
  [\href{http://arxiv.org/abs/1406.2698}{{\tt 1406.2698}}].

\bibitem{Krnjaic:2015mbs}
G.~Krnjaic, \href{http://dx.doi.org/10.1103/PhysRevD.94.073009}{{\it {Probing
  Light Thermal Dark-Matter With a Higgs Portal Mediator}}, } {\em Phys. Rev.}
  {\bf D94} (2016), no.~7 073009, [\href{http://arxiv.org/abs/1512.04119}{{\tt
  1512.04119}}].

\bibitem{lit:b2tip}
P.~Urquijo et~al., {\it {Belle II Physics book}},  {\em in preparation} (2017).
  {\url{https://confluence.desy.de/x/aRoWAg/} (accessed online 08-17-2017).}

\bibitem{Aubert:2008as}
{\bf BaBar Collaboration}, B.~Aubert et~al.,
  \href{http://www-public.slac.stanford.edu/sciDoc/docMeta.aspx?slacPubNumber=slac-pub-13328}{{\it
  {Search for Invisible Decays of a Light Scalar in Radiative Transitions
  $\Upsilon_{3S} \to \gamma A^0$}}, } in {\em {Proceedings of the 34th
  International Conference on High Energy Physics}}, 2008.
\newblock \href{http://arxiv.org/abs/0808.0017}{{\tt 0808.0017}}.

\bibitem{delAmoSanchez:2010ac}
{\bf BaBar Collaboration}, P.~del Amo~Sanchez et~al.,
  \href{http://dx.doi.org/10.1103/PhysRevLett.107.021804}{{\it {Search for
  Production of Invisible Final States in Single-Photon Decays of
  $\Upsilon(1S)$}}, } {\em Phys. Rev. Lett.} {\bf 107} (2011) 021804,
  [\href{http://arxiv.org/abs/1007.4646}{{\tt 1007.4646}}].

\bibitem{Lees:2017lec}
{\bf BaBar Collaboration}, J.~P. Lees et~al., {\it {Search for invisible decays
  of a dark photon produced in e$^+$e$^-$ collisions at BaBar}},
  \href{http://arxiv.org/abs/1702.03327}{{\tt 1702.03327}}.

\bibitem{Alekhin:2015byh}
S.~Alekhin et~al., \href{http://dx.doi.org/10.1088/0034-4885/79/12/124201}{{\it
  {A facility to Search for Hidden Particles at the CERN SPS: the SHiP physics
  case}}, } {\em Rept. Prog. Phys.} {\bf 79} (2016), no.~12 124201,
  [\href{http://arxiv.org/abs/1504.04855}{{\tt 1504.04855}}].

\bibitem{Dror:2017nsg}
J.~A. Dror, R.~Lasenby, and M.~Pospelov, {\it {Dark forces coupled to
  non-conserved currents}},  \href{http://arxiv.org/abs/1707.01503}{{\tt
  1707.01503}}.

\bibitem{Jaeckel:2017tud}
J.~Jaeckel, P.~C. Malta, and J.~Redondo, {\it {Decay photons from the ALP burst
  of type-II supernovae}},  \href{http://arxiv.org/abs/1702.02964}{{\tt
  1702.02964}}.
  
\bibitem{Knapen:2017ebd}
S.~Knapen, T.~Lin, H.~K.~Lou, and T.~Melia,
  {{\it {LHC limits on axion-like particles from heavy-ion collisions}},} in {\em {Proceedings of the Photon 2017 Conference}}, 2017. \newblock \href{http://arxiv.org/abs/1709.07110}{{\tt 1709.07110}}.
  
\bibitem{Hearty:1989pq}
C.~Hearty et~al., \href{http://dx.doi.org/10.1103/PhysRevD.39.3207}{{\it
  {Search for the Anomalous Production of Single Photons in $e^+ e^-$
  Annihilation at $\sqrt{s}=29$-{GeV}}}, } {\em Phys. Rev.} {\bf D39} (1989)
  3207.

\bibitem{Fox:2011fx}
P.~J. Fox, R.~Harnik, J.~Kopp, and Y.~Tsai,
  \href{http://dx.doi.org/10.1103/PhysRevD.84.014028}{{\it {LEP Shines Light on
  Dark Matter}}, } {\em Phys. Rev.} {\bf D84} (2011) 014028,
  [\href{http://arxiv.org/abs/1103.0240}{{\tt 1103.0240}}].

\bibitem{Abdallah:2008aa}
{\bf DELPHI Collaboration}, J.~Abdallah et~al.,
  \href{http://dx.doi.org/10.1140/epjc/s10052-009-0874-9}{{\it {Search for one
  large extra dimension with the DELPHI detector at LEP}}, } {\em Eur. Phys.
  J.} {\bf C60} (2009) 17--23, [\href{http://arxiv.org/abs/0901.4486}{{\tt
  0901.4486}}].

\bibitem{Aarnio:1990vx}
{\bf DELPHI Collaboration}, P.~A. Aarnio et~al.,
  \href{http://dx.doi.org/10.1016/0168-9002(91)90793-P}{{\it {The DELPHI
  detector at LEP}}, } {\em Nucl. Instrum. Meth.} {\bf A303} (1991) 233--276.

\bibitem{Aaboud:2017dor}
{\bf ATLAS Collaboration}, M.~Aaboud et~al., {\it {Search for dark matter at
  $\sqrt{s}=13$ TeV in final states containing an energetic photon and large
  missing transverse momentum with the ATLAS detector}},
  \href{http://arxiv.org/abs/1704.03848}{{\tt 1704.03848}}.

\bibitem{CMS:2017ysu}
{\bf CMS Collaboration}, {\it {Search for Dark Matter Produced in Association
  with a Higgs Boson Decaying to Two Photons}},  2017.
\newblock CMS-PAS-EXO-16-054.

\bibitem{Antreasyan:1990cf}
{\bf Crystal Ball Collaboration}, D.~Antreasyan et~al.,
  \href{http://dx.doi.org/10.1016/0370-2693(90)90254-4}{{\it {Limits on axion
  and light Higgs boson production in $\Upsilon(1S)$ decays}}, } {\em Phys.
  Lett.} {\bf B251} (1990) 204--210.

\bibitem{Aubert:2001tu}
{\bf BaBar Collaboration}, B.~Aubert et~al.,
  \href{http://dx.doi.org/10.1016/S0168-9002(01)02012-5}{{\it {The BaBar
  detector}}, } {\em Nucl. Instrum. Meth.} {\bf A479} (2002) 1--116,
  [\href{http://arxiv.org/abs/hep-ex/0105044}{{\tt hep-ex/0105044}}].

\bibitem{Acciarri:1997im}
{\bf L3 Collaboration}, M.~Acciarri et~al.,
  \href{http://dx.doi.org/10.1016/S0370-2693(97)01003-4}{{\it {Search for new
  physics in energetic single photon production in $e^{+} e^{-}$ annihilation
  at the $Z$ resonance}}, } {\em Phys. Lett.} {\bf B412} (1997) 201--209.

\bibitem{L3:1989aa}
{\bf L3 Collaboration},
  \href{http://dx.doi.org/10.1016/0168-9002(90)90250-A}{{\it {The Construction
  of the L3 Experiment}}, } {\em Nucl. Instrum. Meth.} {\bf A289} (1990)
  35--102.

\bibitem{Chala:2015cev}
M.~Chala, M.~Duerr, F.~Kahlhoefer, and K.~Schmidt-Hoberg,
  \href{http://dx.doi.org/10.1016/j.physletb.2016.02.006}{{\it {Tricking
  Landau-Yang: How to obtain the diphoton excess from a vector resonance}}, }
  {\em Phys. Lett.} {\bf B755} (2016) 145--149,
  [\href{http://arxiv.org/abs/1512.06833}{{\tt 1512.06833}}].

\bibitem{Aaltonen:2013mfa}
{\bf CDF Collaboration}, T.~A. Aaltonen et~al.,
  \href{http://dx.doi.org/10.1103/PhysRevLett.112.111803}{{\it {First Search
  for Exotic Z Boson Decays into Photons and Neutral Pions in Hadron
  Collisions}}, } {\em Phys. Rev. Lett.} {\bf 112} (2014) 111803,
  [\href{http://arxiv.org/abs/1311.3282}{{\tt 1311.3282}}].

\bibitem{Aad:2015bua}
{\bf ATLAS Collaboration}, G.~Aad et~al.,
  \href{http://dx.doi.org/10.1140/epjc/s10052-016-4034-8}{{\it {Search for new
  phenomena in events with at least three photons collected in $pp$ collisions
  at $\sqrt{s}$ = 8 TeV with the ATLAS detector}}, } {\em Eur. Phys. J.} {\bf
  C76} (2016), no.~4 210, [\href{http://arxiv.org/abs/1509.05051}{{\tt
  1509.05051}}].

\bibitem{Tsai:1986tx}
Y.-S. Tsai, \href{http://dx.doi.org/10.1103/PhysRevD.34.1326}{{\it {Axion
  bremsstrahlung by an electron beam}}, } {\em Phys. Rev.} {\bf D34} (1986)
  1326.

\bibitem{Bjorken:2009mm}
J.~D. Bjorken, R.~Essig, P.~Schuster, and N.~Toro,
  \href{http://dx.doi.org/10.1103/PhysRevD.80.075018}{{\it {New Fixed-Target
  Experiments to Search for Dark Gauge Forces}}, } {\em Phys. Rev.} {\bf D80}
  (2009) 075018, [\href{http://arxiv.org/abs/0906.0580}{{\tt 0906.0580}}].

\bibitem{Riordan:1987aw}
E.~M. Riordan et~al., \href{http://dx.doi.org/10.1103/PhysRevLett.59.755}{{\it
  {A Search for Short Lived Axions in an Electron Beam Dump Experiment}}, }
  {\em Phys. Rev. Lett.} {\bf 59} (1987) 755.

\bibitem{Krasny}
M.~W. Krasny et~al., {\it {Recent searches for short-lived pseudoscalar bosons
  in electron beam-dump experiments}},  in {\em {Proceedings of the EPS
  conference}}, 1987.

\bibitem{Dobrich:2017gcm}
B.~D{\"o}brich, {\it {Axion-like Particles from Primakov production in
  beam-dumps}},  2017.
\newblock \href{http://arxiv.org/abs/1708.05776}{{\tt 1708.05776}}.

\bibitem{Bjorken:1988as}
J.~D. Bjorken, S.~Ecklund, W.~R. Nelson, A.~Abashian, C.~Church, et~al.,
  \href{http://dx.doi.org/10.1103/PhysRevD.38.3375}{{\it {Search for Neutral
  Metastable Penetrating Particles Produced in the SLAC Beam Dump}}, } {\em
  Phys. Rev.} {\bf D38} (1988) 3375.

\bibitem{Bergsma:1985qz}
{\bf CHARM Collaboration}, F.~Bergsma et~al.,
  \href{http://dx.doi.org/10.1016/0370-2693(85)90400-9}{{\it {Search for Axion
  Like Particle Production in 400-{GeV} Proton - Copper Interactions}}, } {\em
  Phys. Lett.} {\bf B157} (1985) 458--462.

\bibitem{Blumlein:1990ay}
J.~Blumlein et~al., \href{http://dx.doi.org/10.1007/BF01548556}{{\it {Limits on
  neutral light scalar and pseudoscalar particles in a proton beam dump
  experiment}}, } {\em Z. Phys.} {\bf C51} (1991) 341--350.
  
\bibitem{Blumlein:1991xh}
J.~Blumlein et~al., \href{http://dx.doi.org/10.1142/S0217751X9200171X}{{\it {Limits on the mass of light (pseudo)scalar particles from Bethe-Heitler e+ e- and mu+ mu- pair production in a proton - iron beam dump experiment}}, } Int.\ J.\ Mod.\ Phys.\ A {\bf 7} (1992) 3835.

\bibitem{Payez:2014xsa}
A.~Payez, C.~Evoli, T.~Fischer, M.~Giannotti, A.~Mirizzi, et~al.,
  \href{http://dx.doi.org/10.1088/1475-7516/2015/02/006}{{\it {Revisiting the
  SN1987A gamma-ray limit on ultralight axion-like particles}}, } {\em JCAP}
  {\bf 1502} (2015), no.~02 006, [\href{http://arxiv.org/abs/1410.3747}{{\tt
  1410.3747}}].

\bibitem{Brockway:1996yr}
J.~W. Brockway, E.~D. Carlson, and G.~G. Raffelt,
  \href{http://dx.doi.org/10.1016/0370-2693(96)00778-2}{{\it {SN1987A gamma-ray
  limits on the conversion of pseudoscalars}}, } {\em Phys. Lett.} {\bf B383}
  (1996) 439--443, [\href{http://arxiv.org/abs/astro-ph/9605197}{{\tt
  astro-ph/9605197}}].

\bibitem{Raffelt:1996wa}
G.~G. Raffelt, {\em {Stars as laboratories for fundamental physics}}.
\newblock Chicago, USA: Univ. Pr., 1996.
\newblock \url{http://wwwth.mpp.mpg.de/members/raffelt/mypapers/199613.pdf}.

\bibitem{Raffelt:1988rx}
G.~G. Raffelt and G.~D. Starkman,
  \href{http://dx.doi.org/10.1103/PhysRevD.40.942}{{\it {Stellar energy
  transfer by keV mass scalars}}, } {\em Phys. Rev.} {\bf D40} (1989) 942.

\bibitem{Janka:2012wk}
H.-T. Janka, \href{http://dx.doi.org/10.1146/annurev-nucl-102711-094901}{{\it
  {Explosion Mechanisms of Core-Collapse Supernovae}}, } {\em Ann. Rev. Nucl.
  Part. Sci.} {\bf 62} (2012) 407--451,
  [\href{http://arxiv.org/abs/1206.2503}{{\tt 1206.2503}}].

\bibitem{Raffelt:1987yb}
G.~G. Raffelt and D.~S.~P. Dearborn,
  \href{http://dx.doi.org/10.1103/PhysRevD.37.549}{{\it {Bounds on Weakly
  Interacting Particles From Observational Lifetimes of Helium Burning Stars}},
  } {\em Phys. Rev.} {\bf D37} (1988) 549--551.

\bibitem{Buras:2005rp}
R.~Buras, M.~Rampp, H.~T. Janka, and K.~Kifonidis,
  \href{http://dx.doi.org/10.1051/0004-6361:20053783}{{\it {Two-dimensional
  hydrodynamic core-collapse supernova simulations with spectral neutrino
  transport. 1. Numerical method and results for a 15 solar mass star}}, } {\em
  Astron. Astrophys.} {\bf 447} (2006) 1049--1092,
  [\href{http://arxiv.org/abs/astro-ph/0507135}{{\tt astro-ph/0507135}}].

\bibitem{Fischer:2016cyd}
T.~Fischer, S.~Chakraborty, M.~Giannotti, A.~Mirizzi, A.~Payez, et~al.,
  \href{http://dx.doi.org/10.1103/PhysRevD.94.085012}{{\it {Probing axions with
  the neutrino signal from the next galactic supernova}}, } {\em Phys. Rev.}
  {\bf D94} (2016), no.~8 085012, [\href{http://arxiv.org/abs/1605.08780}{{\tt
  1605.08780}}].

\bibitem{Friedland:2012hj}
A.~Friedland, M.~Giannotti, and M.~Wise,
  \href{http://dx.doi.org/10.1103/PhysRevLett.110.061101}{{\it {Constraining
  the Axion-Photon Coupling with Massive Stars}}, } {\em Phys. Rev. Lett.} {\bf
  110} (2013), no.~6 061101, [\href{http://arxiv.org/abs/1210.1271}{{\tt
  1210.1271}}].

\bibitem{Aaboud:2017bwk}
{\bf ATLAS Collaboration}, M.~Aaboud et~al.,
  \href{http://dx.doi.org/doi:10.1038/nphys4208}{{\it {Evidence for light-by-light scattering in heavy-ion collisions with the ATLAS detector at the LHC}}, } {\em Nature Phys.} {\bf 13} (2017) no.9,  852, [\href{http://arxiv.org/abs/1702.01625}{{\tt 1702.01625}}].
  
\bibitem{Ackermann:2015lka}
{\bf Fermi-LAT Collaboration}, M.~Ackermann et~al.,
  \href{http://dx.doi.org/10.1103/PhysRevD.91.122002}{{\it {Updated search for
  spectral lines from Galactic dark matter interactions with pass 8 data from
  the Fermi Large Area Telescope}}, } {\em Phys. Rev.} {\bf D91} (2015), no.~12
  122002, [\href{http://arxiv.org/abs/1506.00013}{{\tt 1506.00013}}].

\bibitem{Albert:2014hwa}
{\bf Fermi-LAT Collaboration}, A.~Albert, G.~A. Gomez-Vargas, M.~Grefe,
  C.~Munoz, C.~Weniger, et~al.,
  \href{http://dx.doi.org/10.1088/1475-7516/2014/10/023}{{\it {Search for
  100\,MeV to 10\,GeV $\gamma$-ray lines in the Fermi-LAT data and implications
  for gravitino dark matter in $\mu\nu$SSM}}, } {\em JCAP} {\bf 1410} (2014),
  no.~10 023, [\href{http://arxiv.org/abs/1406.3430}{{\tt 1406.3430}}].

\bibitem{Frandsen:2012db}
M.~T. Frandsen, U.~Haisch, F.~Kahlhoefer, P.~Mertsch, and K.~Schmidt-Hoberg,
  \href{http://dx.doi.org/10.1088/1475-7516/2012/10/033}{{\it {Loop-induced
  dark matter direct detection signals from gamma-ray lines}}, } {\em JCAP}
  {\bf 1210} (2012) 033, [\href{http://arxiv.org/abs/1207.3971}{{\tt
  1207.3971}}].

\bibitem{Gondolo:1990dk}
P.~Gondolo and G.~Gelmini,
  \href{http://dx.doi.org/10.1016/0550-3213(91)90438-4}{{\it {Cosmic abundances
  of stable particles: Improved analysis}}, } {\em Nucl. Phys.} {\bf B360}
  (1991) 145--179.

\bibitem{Belanger:2014vza}
G.~B\'elanger, F.~Boudjema, A.~Pukhov, and A.~Semenov,
  \href{http://dx.doi.org/10.1016/j.cpc.2015.03.003}{{\it {micrOMEGAs4.1: two
  dark matter candidates}}, } {\em Comput. Phys. Commun.} {\bf 192} (2015)
  322--329, [\href{http://arxiv.org/abs/1407.6129}{{\tt 1407.6129}}].

\bibitem{Binder:2017rgn}
T.~Binder, T.~Bringmann, M.~Gustafsson, and A.~Hryczuk, {\it {Early kinetic
  decoupling of dark matter: when the standard way of calculating the thermal
  relic density fails}},  \href{http://arxiv.org/abs/1706.07433}{{\tt
  1706.07433}}.

\bibitem{Alwall:2014hca}
J.~Alwall et~al., \href{http://dx.doi.org/10.1007/JHEP07(2014)079}{{\it {The
  automated computation of tree-level and next-to-leading order differential
  cross sections, and their matching to parton shower simulations}}, } {\em
  JHEP} {\bf 07} (2014) 079, [\href{http://arxiv.org/abs/1405.0301}{{\tt
  1405.0301}}].

\bibitem{CarloniCalame:2003yt}
C.~Calame et~al.,
  \href{http://dx.doi.org/10.1016/j.nuclphysbps.2004.02.008}{{\it {The BABAYAGA
  event generator}}, } {\em Nucl. Phys. Proc. Suppl.} {\bf 131} (2004) 48--55,
  [\href{http://arxiv.org/abs/hep-ph/0312014}{{\tt hep-ph/0312014}}].
  [,48(2003)].

\bibitem{CarloniCalame:2001ny}
C.~Calame, \href{http://dx.doi.org/10.1016/S0370-2693(01)01108-X}{{\it {An
  Improved parton shower algorithm in QED}}, } {\em Phys. Lett.} {\bf B520}
  (2001) 16--24, [\href{http://arxiv.org/abs/hep-ph/0103117}{{\tt
  hep-ph/0103117}}].

\bibitem{CarloniCalame:2000pz}
C.~Calame et~al., \href{http://dx.doi.org/10.1016/S0550-3213(00)00356-4}{{\it
  {Large angle Bhabha scattering and luminosity at flavor factories}}, } {\em
  Nucl. Phys.} {\bf B584} (2000) 459--479,
  [\href{http://arxiv.org/abs/hep-ph/0003268}{{\tt hep-ph/0003268}}].

\bibitem{Aulchenko:2017lmh}
V.~Aulchenko, A.~Bobrov, T.~Ferber, A.~Kuzmin, K.~Miyabayshi, et~al.,
  \href{http://dx.doi.org/10.1088/1748-0221/12/08/C08001}{{\it {Time and energy
  reconstruction at the electromagnetic calorimeter of the Belle-II detector}},
  } {\em JINST} {\bf 12} (2017), no.~08 C08001.

\bibitem{Czyz:2017veo}
H.~Czyz, P.~Kisza and S.~Tracz, {\it {Modeling interactions of photons with pseudoscalar and vector mesons}},  \href{http://arxiv.org/abs/1711.00820}{{\tt
  1711.00820}}.
  
\bibitem{Abashian:2000cg}
{\bf Belle Collaboration}, A.~Abashian et~al.,
  \href{http://dx.doi.org/10.1016/S0168-9002(01)02013-7}{{\it {The Belle
  Detector}}, } {\em Nucl. Instrum. Meth.} {\bf A479} (2002) 117--232.

\end{thebibliography}
\end{document}